\documentclass[11pt,tightenlines,eqsecnum,floats,aps,amssymb,nofootinbib,prd,shownopacs,floatfix,superscriptaddress]{revtex4-2}

\usepackage{braket}
\usepackage[dvipsnames]{xcolor}
\usepackage{bm}
\usepackage{dirtytalk}
\usepackage{hyperref}
\usepackage{graphicx}
\usepackage{comment}
\usepackage{ragged2e}
\usepackage{etoolbox}
\usepackage{relsize}
\usepackage{float}
\usepackage{amsmath}
\usepackage{amssymb}
\usepackage{mathrsfs}
\usepackage{parskip}
\usepackage[T1]{fontenc}
\usepackage{libertinus}
\usepackage{layouts}
\usepackage[symbol]{footmisc}
\usepackage{subcaption}

\usepackage{soul}

\usepackage[colorinlistoftodos,prependcaption,textsize=tiny]{todonotes}

\makeatletter
\newrobustcmd{\fixappendix}{%
 \patchcmd{\l@section}{1.5em}{7em}{}{}%
 \patchcmd{\l@subsection}{2.3em}{7em}{}{}%
}
\makeatother

\hypersetup{
  colorlinks=true,
  linkcolor=blue,
  citecolor=blue,
  filecolor=magenta,   
  urlcolor=blue,
  pdfpagemode=FullScreen,
}

\newcommand{\dd}{\ensuremath{\mathrm{d}}}

\renewcommand{\footnoterule}{%
 \kern -3pt
 \hrule width 0.4\columnwidth
 \kern 2.6pt
}

\begin{document}

\begin{abstract}

How many canonical degrees of freedom are dynamically needed to reproduce the Hamiltonian evolution of a regulated field theory? We address this question for a UV/IR-regularised classical scalar field by identifying the minimal symplectic dimension required to reproduce a single trajectory by an autonomous Hamiltonian system. Using symplectic model order reduction as a structure-preserving diagnostic, we show that, for the free scalar field, this dimension is controlled not by the volume-extensive number of regulated field modes, but by the smaller number of distinct normal-mode frequencies below the ultraviolet cutoff. In a flat cubic box this gives an area-type scaling with the size of the region, up to slowly varying corrections. For geodesic balls in maximally symmetric curved spaces, positive curvature produces a super-area enhancement that remains subextensive, while negative curvature suppresses the scaling, with the flat result recovered smoothly in the small-curvature limit. Numerical experiments indicate that the same frequency-based mechanism persists in weakly interacting \(\lambda\phi^4\) theory on quasi-integrable time scales. Beyond the counting result, the reduced dynamics has a characteristic algebraic structure: it decomposes into independent oscillator blocks, while reconstructed field modes are linear combinations of these blocks and have Poisson brackets governed by a projector rather than the identity. In this precise classical sense, overlapping degrees of freedom arise dynamically, without modifying the canonical structure by hand. The results provide a controlled setting in which area-type dynamical scaling and overlap structures can be studied before quantisation, and help separate ordinary Hamiltonian compression effects from genuinely gravitational mechanisms.

\end{abstract}

\title{Area Scaling of Dynamical Degrees of Freedom in Regularised Scalar Field Theory}
\author{Oliver Friedrich}
\email{oliver.friedrich@lmu.de}
\affiliation{University Observatory, Faculty of Physics, Ludwig-Maximilians-Universität, Scheinerstraße 1, 81677 Munich, Germany}
\affiliation{Excellence Cluster ORIGINS, Boltzmannstr.\ 2, 85748 Garching, Germany}

\author{Kristina Giesel}
\email{kristina.giesel@fau.de}
\affiliation{Institute for Quantum Gravity, Theoretical Physics III, Department of Physics, Friedrich-Alexander-Universit\"at Erlangen-N\"urnberg, Staudtstr. 7, 91058 Erlangen, Germany.}

\author{Varun Kushwaha$^*$}
\email{varun.kushwaha@lmu.de}
\thanks{corresponding author}
\affiliation{University Observatory, Faculty of Physics, Ludwig-Maximilians-Universität, Scheinerstraße 1, 81677 Munich, Germany}

\maketitle

\clearpage
\section*{Abbreviations and notation}

\begingroup
\setlength{\parindent}{0pt}
\setlength{\parskip}{0pt}
\renewcommand{\arraystretch}{1.2}

\subsection*{Abbreviations}

\begin{tabular}{p{0.17\textwidth}p{0.78\textwidth}}
SMOR  & symplectic model order reduction \\
ROM  & reduced–order model \\
SVD  & singular value decomposition \\
cSVD  & complex singular value decomposition \\
PSD  & proper symplectic decomposition \\
SMG  & symplectic manifold Galerkin (projection)
\end{tabular}

\vspace{1.5em}

\subsection*{Notation}

\begin{tabular}{p{0.17\textwidth}p{0.78\textwidth}}
$\mathcal B$ 
 & spatial region (flat box or geodesic ball) \\

$L_{\mathrm{IR}}$ 
 & infrared box size / characteristic linear size of $\mathcal B$ \\

$\Lambda_{\mathrm{UV}}$ 
 & ultraviolet cutoff in momentum space \\

$N$ 
 & number of discretised field values (configuration DoF) \\

$\bm z=(\bm Q,\bm P)^\top$ 
 & full phase–space coordinate in $\mathbb R^{2N}$ \\

$\widetilde{\bm z}\in\mathbb R^{2m}$ 
 & reduced phase–space coordinate (ROM canonical variables) \\

$\mathbb J_{2k}$ 
 & standard symplectic matrix in dimension $2k$ \\

$\bm V\in\mathbb R^{2N\times 2m}$ 
 & symplectic embedding matrix, $\ \bm V^\top\mathbb J_{2N}\bm V=\mathbb J_{2m}$ \\

$\bm V^{+}$ 
 & symplectic adjoint of $\bm V$, $\ \bm V^{+}=\mathbb J_{2m}^\top\,\bm V^\top\,\mathbb J_{2N}$ \\

$\mathbb X_s$ 
 & real snapshot matrix of sampled trajectories, 
  $\displaystyle 
  \mathbb X_s=
  \begin{pmatrix}
   \mathbb X_s^{(Q)} \\[2pt]
   \mathbb X_s^{(P)}
  \end{pmatrix}
  \in\mathbb R^{2N\times M}$ \\

$\mathbb X_s^{(Q)}$, $\mathbb X_s^{(P)}$ 
 & blocks of $\mathbb X_s$ collecting $Q(t_j)$ and $P(t_j)$ respectively, 
  each in $\mathbb R^{N\times M}$ \\

$\mathbb X_c$ 
 & complexified snapshot matrix, 
  $\displaystyle 
   \mathbb X_c := \mathbb X^{(Q)} + i\,\mathbb X^{(P)} \in\mathbb C^{N\times M}$ \\

$\mathcal E_{\mathrm{proj}}$ 
 & relative projection error of a candidate reduced subspace \\

$n_\Omega$ 
 & number of distinct physical normal–mode frequencies below $\Lambda_{\mathrm{UV}}$ \\

$\Omega_s$ 
 & distinct normal–mode frequency associated with shell $s$ \\

$\Upsilon$ 
 & partial isometry mapping reduced oscillators to apparent modes \\

$\mathcal C=\Upsilon\Upsilon^\dagger$ 
 & overlap projector governing Poisson brackets / commutators of apparent modes \\
\end{tabular}

\endgroup

\newpage
{\small\tableofcontents}
\newpage
\markboth{Area Scaling of Dynamical Degrees of Freedom in Regulated Scalar Fields}{}

\section{Introduction}
\label{sec:introduction}

Low-energy physics is usually organised in the language of effective field theory. If a theory of quantum gravity has a semiclassical regime, then in that regime we expect to recover, at least approximately, a spacetime geometry together with quantum fields propagating on it. This effective description is local: after a spatial region and a UV cutoff have been specified, the field is represented by a finite set of regulated degrees of freedom distributed throughout the region. For a scalar field in three spatial dimensions, the number of such kinematical degrees of freedom grows with the volume of the region \cite{PadmanabhanBook,Peskin:1995ev}.

Gravity suggests that this volume-extensive kinematical count should not be identified too quickly with a count of independent microscopic resources. Black-hole entropy, entropy bounds and holographic arguments indicate that the independent information associated with a gravitating region is controlled by an area rather than by the volume suggested by naive local effective-field-theory counting \cite{'tHooft:1993gx,Susskind1995,Bousso2002,Bousso1999b,Bousso1999a,Bousso_2014,Fischler1998}. These arguments do not tell us, by themselves, how the effective field-theory variables should be related to the independent variables of the underlying theory. They do, however, motivate a precise distinction: the local degrees of freedom used in an effective description may be apparent degrees of freedom, while the independent physical resources may be fewer.

A useful way to express this distinction is through overlap. If many apparent degrees of freedom are represented using fewer underlying ones, then the apparent variables cannot all remain exactly independent on the full microscopic state space. In quantum language, this means that the apparent qubits or local algebras cannot be exact independent tensor factors globally. They may nevertheless behave as approximately independent degrees of freedom within a suitable sector of states, much as logical operators in a code subspace reproduce the algebra of bulk effective fields only on that subspace \cite{Akers_2022,akers2022blackholeinteriornonisometric,Akers_2021}. In this sense, overlap is not merely an imperfection of a model; it is the algebraic price of representing many apparent local variables using fewer underlying resources.

This point is made concrete in constructions of overlapping qubits and non-isometric maps, where approximately local effective observables are represented using a smaller Hilbert space \cite{Cao_2025}. It also appears in the overlapping-mode field-theory model of Ref.~\cite{Friedrich2024}, where many apparent field modes are built from a smaller set of underlying degrees of freedom, leading to an effective area-type count. In these constructions the existence of a smaller underlying system is part of the starting point. The compression map, and hence the pattern of overlap, is introduced as an input. This leaves open a natural question: can an overlap structure be selected by the dynamics of an ordinary field theory itself, rather than postulated as an external map?

The present paper addresses this question in a controlled setting. We start with an ordinary Hamiltonian scalar field on a fixed background, impose UV and IR regulators, and ask how many canonical directions are dynamically required to reproduce its motion. We do not ask whether the full regulated field theory can be replaced by fewer variables for all possible initial data. It cannot: arbitrary initial data require the full regulated phase space. Instead we ask a sector-wise question, in the simplest possible classical form. Once an initial condition is fixed, the Hamiltonian selects a particular trajectory. The question is whether that trajectory can be reproduced by a smaller autonomous Hamiltonian system, and if so, how many canonical degrees of freedom that smaller system needs.

This trajectory-wise formulation is not meant to be a weakness of the construction. It is the classical analogue of restricting attention to a sector of states or processes rather than demanding an exact representation on the full state space \cite{Akers_2022,akers2022blackholeinteriornonisometric,Akers_2021}. A code-subspace description in quantum gravity is not expected to reproduce all states of the microscopic theory; it is expected to reproduce a controlled family of effective states and observables. Similarly, the trajectory-wise question asks for the degrees of freedom required by a specified dynamical sector of the classical theory. For the free scalar field considered below, the resulting dimension is in fact stable for generic initial data that excite the same set of frequency shells: the orientation of the reduced subspace depends on the initial condition, but the minimal dimension is controlled by the spectrum.

This viewpoint separates two different counts. The regulator tells us how many canonical variables are kinematically available. The Hamiltonian trajectory tells us how many canonical directions are dynamically used. If the trajectory can be reproduced by a smaller autonomous Hamiltonian system, then the original field variables contain redundancy from the point of view of that motion. Reconstructing the original variables from the smaller system then produces an induced overlap algebra. Thus the same question tests both the scaling of dynamically required degrees of freedom and the emergence of non-independent apparent variables.

To formulate the question precisely, one must compare the full regulated phase space with a smaller phase space. Such a comparison necessarily involves a reconstruction map: reduced variables are evolved in the smaller system and then embedded back into the original field variables. A generic reduction of coordinates is not sufficient for our purposes. It may destroy the canonical pairing between coordinates and momenta, and it may fail to define a closed autonomous Hamiltonian system. Since the objects we want to count are canonical degrees of freedom, the reduced description must itself preserve the symplectic structure. We therefore ask: given a trajectory of the UV/IR-regularised scalar field, what is the smallest symplectic phase space on which an autonomous Hamiltonian system can reproduce that trajectory, exactly or within a prescribed tolerance?

The tool we use to answer this question is symplectic model order reduction (SMOR) \cite{peng2015symplecticmodelreductionhamiltonian,hesthaven2021structurepreservingmodelorderreduction,buchfink2019symplecticmodelorderreduction,buchfink2021symplecticmodelreductionhamiltonian}. SMOR constructs reduced phase spaces from trajectory data while preserving the symplectic form. The reduced variables therefore still come in canonical pairs, and the reduced equations remain Hamiltonian. Although SMOR is usually used as a numerical acceleration method, here we use it as a diagnostic of dynamical dimension. For a given trajectory, the minimal reduced symplectic dimension measures the number of canonical directions required to reproduce the motion.

For the free scalar field, this diagnostic becomes analytically transparent. After regularisation, the free Hamiltonian is a finite sum of harmonic normal modes. A trajectory is therefore a superposition of oscillatory components. The minimal symplectic dimension is not controlled by the total number of mode labels below the UV cutoff. It is controlled by the number of distinct normal-mode frequencies present in the trajectory. Modes with the same frequency share the same temporal oscillator structure, and a single reduced canonical block can represent the corresponding shell contribution. Thus, for the free theory, the SMOR problem becomes a degeneracy-stripped frequency-counting problem.

This observation gives the main result. In a flat cubic box, the number of distinct normal-mode frequencies below the UV cutoff grows proportionally to the boundary area, up to slowly varying corrections. On geodesic balls in maximally symmetric curved spaces, positive curvature produces a super-area enhancement that remains subextensive, while negative curvature suppresses the scaling. The flat result is recovered smoothly in the small-curvature limit. We further show numerically that, in weakly interacting $\lambda\phi^4$ theory and on quasi-integrable time scales, the same frequency-based mechanism continues to organise the minimal symplectic dimension; see Fig.~\ref{fig:projection_error}.

The reduced description also gives a concrete algebraic realisation of overlap. The reduced system contains fewer canonical oscillator blocks than the original regulated field theory. When the original field modes are reconstructed from the reduced variables, many apparent full-order modes depend on the same smaller set of canonical directions. Their induced Poisson brackets are therefore governed by a finite-rank projector rather than by the identity. In this precise classical sense, the apparent field modes become overlapping degrees of freedom. The area-type scaling and the overlap algebra are two sides of the same reduction: fewer canonical directions are dynamically needed, and the original mode variables become non-independent when restricted to the reduced phase space.

The lesson is modest but useful. The analysis does not produce a holographic duality, nor does it identify the microscopic degrees of freedom of quantum gravity. It isolates a classical Hamiltonian mechanism by which a volume-extensive set of apparent field variables can reduce, for a controlled dynamical sector, to a smaller set of canonical directions with an induced overlap algebra. 

The remainder of the article is organised as follows. Section~\ref{sec:SMOR} introduces the SMOR framework and defines the minimal symplectic dimension. Section~\ref{sec:application} applies this framework to a UV/IR-regularised scalar field in flat and curved geometries, presenting analytical and numerical results for the scaling of dynamically relevant degrees of freedom. Section~\ref{sec:discussion} discusses the assumptions underlying the analysis and provides a dynamical interpretation in terms of action-angle variables, state dependence, and overlap structures. Section~\ref{sec:Conclusions and Outlook} summarises the conclusions and outlines directions for future work. Technical details are collected in the appendices.

\section{Symplectic model order reduction and dynamical dimension}
\label{sec:SMOR}

We now make the trajectory-wise compression problem precise.  The input is a finite-dimensional Hamiltonian system, obtained below from a UV/IR-regularised scalar field, together with one of its trajectories.  The output we seek is a reduced Hamiltonian system that reproduces this trajectory using as few canonical directions as possible.

The reduction must preserve symplectic structure.  A generic projection may give a lower-dimensional description of the sampled data, but it need not preserve the pairing between coordinates and momenta, and it need not define Hamiltonian equations on the reduced space.  Symplectic model order reduction (SMOR) avoids this problem by restricting to reductions that preserve the symplectic form.  The reduced variables therefore still come in canonical pairs, and the reduced dynamics remain Hamiltonian.

In the main text we keep only the ingredients needed for the analysis: the snapshot matrix, the symplectic projection error, the complex-SVD construction of the reduced basis, the reduced Hamiltonian system, and the resulting definition of the minimal symplectic dimension.  A more geometric review of the underlying SMOR framework is given in Appendix~\ref{app:SMOR-background}; broader introductions can be found in Refs.~\cite{peng2015symplecticmodelreductionhamiltonian,hesthaven2021structurepreservingmodelorderreduction,buchfink2019symplecticmodelorderreduction,buchfink2024modelreductionmanifoldsdifferential}.

After regularisation, the field theory is a finite-dimensional Hamiltonian system with phase-space point
\begin{align}
    \bm z =
    \begin{pmatrix}
        \bm Q \\[2pt] \bm P
    \end{pmatrix}
    \in \mathbb R^{2N},
    \qquad
    \mathbb J_{2N} =
    \begin{pmatrix}
        0 & \mathbb I_N \\
        -\mathbb I_N & 0
    \end{pmatrix}.
\end{align}
Here $N$ is the number of discretized configuration-space field variables, so the full phase space has dimension $2N$. The vectors $\bm{Q}$ and $\bm{P}$ collect the corresponding canonical coordinates and momenta, and $\mathbb{J}_{2N}$ denotes the standard symplectic matrix in dimension $2N$. The full equations of motion are
\begin{align}
    \dot{\bm z}
    =
    X_{\mathcal H}(\bm z)
    =
    \mathbb J_{2N}\nabla\mathcal H(\bm z),
    \label{eq:full-hamiltonian-system}
\end{align}
where $\mathcal H$ is the regulated Hamiltonian.  A linear reduced phase space of symplectic dimension $2m$ is specified by a matrix
\begin{align}
    \bm V\in\mathbb R^{2N\times 2m},
    \qquad
    \bm V^\top\mathbb J_{2N}\bm V
    =
    \mathbb J_{2m}.
    \label{eq:symplectic-basis-condition}
\end{align}
This condition says that the columns of $\bm V$ span a symplectic subspace of the full phase space.  Reduced coordinates $\widetilde{\bm z}\in\mathbb R^{2m}$ are embedded into the full phase space by
\begin{align}
    \bm z
    \approx
    \bm V\widetilde{\bm z}.
\end{align}
The corresponding symplectic adjoint is
\begin{align}
    \bm V^+
    :=
    \mathbb J_{2m}^{\top}\bm V^\top\mathbb J_{2N},
    \label{eq:symplectic-adjoint}
\end{align}
which satisfies $\bm V^+\bm V=\mathbb I_{2m}$.  It defines the symplectic projector
\begin{align}
    P_{\bm V}
    :=
    \bm V\bm V^+
    \label{eq:symplectic-projector}
\end{align}
onto the embedded reduced space.

\subsection{Snapshot matrix and symplectic projection error}
\label{sec:snapshot-projection-error}

We apply this construction to individual Hamiltonian trajectories.  Fix the physical parameters and an initial condition $\bm z_0$.  We denote by $\Phi^t$ the Hamiltonian flow generated by \eqref{eq:full-hamiltonian-system}, so that the full solution is
\begin{align}
    \bm z(t)
    =
    \Phi^t(\bm z_0).
\end{align}
Choose sampling times $t_j$ in an observation window $I=[t_0,t_M]$, with $j=1,\ldots,M$.  The sampling need not be uniform for the definitions below; in the numerical applications we use sufficiently fine sampling over a long enough window to resolve the relevant temporal frequencies.  The sampled trajectory is assembled into the snapshot matrix
\begin{align}
    \mathbb X_s
    =
    \begin{pmatrix}
        \mathbb X_s^{(Q)} \\[2pt]
        \mathbb X_s^{(P)}
    \end{pmatrix}
    =
    \begin{pmatrix}
        \cdots & \bm Q(t_j) & \cdots \\
        \cdots & \bm P(t_j) & \cdots
    \end{pmatrix}
    \in \mathbb R^{2N\times M}.
    \label{eq:snapshot-matrix}
\end{align}
Each column is one full phase-space state along the orbit.  The snapshot matrix is therefore the finite data object encoding the part of phase space explored by the trajectory during the observation window.

Given a candidate symplectic basis $\bm V$, the part of the sampled trajectory not captured by the reduced symplectic subspace is measured by the Frobenius projection error
\begin{align}
    \big\|
        (\mathbf I-P_{\bm V})\mathbb X_s
    \big\|_F
    =
    \big\|
        (\mathbf I-\bm V\bm V^+)\mathbb X_s
    \big\|_F .
    \label{eq:proj-loss}
\end{align}
This quantity is small when all sampled states lie close to $\operatorname{range}(\bm V)$, and it vanishes when the sampled trajectory is contained in that subspace.

In many model-reduction applications, one constructs a reduced space for a family of solutions obtained by varying parameters or initial data \cite{hesthaven2021structurepreservingmodelorderreduction,buchfink2024modelreductionmanifoldsdifferential}.  Such a space measures the collective complexity of that family.  Here we use SMOR differently.  We ask how many canonical directions are required by one orbit of the regulated field theory.  This trajectory-wise question isolates the dynamical dimension actually used by Hamiltonian evolution, rather than the kinematical dimension made available by the regulator.  Maximising this number over admissible initial data then gives the largest dynamically relevant canonical dimension associated with the chosen regulator and geometry.

\subsection{Complex SVD and the PSD basis}
\label{sec:csvd-psd-basis}

The snapshot matrix tells us which region of phase space the trajectory visits.  To decide whether this information can be represented by a smaller Hamiltonian system, we must ask how small the projection error \eqref{eq:proj-loss} can be made for a $2m$-dimensional symplectic subspace.  In other words, we must choose the reduced symplectic basis from the data.

For general Hamiltonian systems, globally optimal proper symplectic decomposition (PSD) bases are not known in closed form.  We therefore restrict to the class of \emph{ortho-symplectic} matrices,
\begin{align}
    \mathbb S(2m,2N)
    :=
    \left\{
        \bm V\in\mathbb R^{2N\times 2m}
        \ \middle|\
        \bm V^\top\mathbb J_{2N}\bm V=\mathbb J_{2m},
        \quad
        \bm V^\top\bm V=\mathbb I_{2m}
    \right\}.
    \label{eq:ortho-symplectic-set}
\end{align}
The symplectic condition preserves the canonical structure, while the additional orthonormality condition makes the minimisation problem explicitly solvable.  This is also the class naturally produced by the complex-SVD construction used below.  Its role in defining the reduced Hamiltonian dynamics is made explicit in Sec.~\ref{sec:reduction-step}.

The construction exploits the canonical pairing of the variables.  Rather than treating $\bm Q$ and $\bm P$ as unrelated real data, we combine each canonical pair into a complex variable and define
\begin{align}
    \mathbb X_c
    :=
    \mathbb X_s^{(Q)}
    +
    i\,\mathbb X_s^{(P)}
    \in
    \mathbb C^{N\times M}.
    \label{eq:complex-snapshot-matrix}
\end{align}
Let
\begin{align}
    \mathbb X_c
    =
    \mathbb U\,\bm\Sigma\,\mathbb W^\ast
\end{align}
be its singular value decomposition, and let $\mathbb U_m\in\mathbb C^{N\times m}$ denote the matrix formed from the $m$ dominant left singular vectors.  Writing
\begin{align}
    \mathbb U_m
    =
    \bm\Phi+i\bm\Psi,
\end{align}
with $\bm\Phi,\bm\Psi\in\mathbb R^{N\times m}$, we obtain the real phase-space basis
\begin{align}
    \bm V^{(\rm opt)}_m
    =
    \begin{pmatrix}
        \bm\Phi & -\bm\Psi \\
        \bm\Psi & \ \bm\Phi
    \end{pmatrix}
    \in
    \mathbb R^{2N\times 2m}.
    \label{eq:Vstar-cSVD}
\end{align}
This is the key point of the complex-SVD construction: $\bm V^{(\rm opt)}_m$ is not merely an orthonormal basis for a real $2m$-dimensional subspace; it is an ortho-symplectic basis,
\begin{align}
    (\bm V^{(\rm opt)}_m)^\top \bm V^{(\rm opt)}_m
    =
    \mathbb I_{2m},
    \qquad
    (\bm V^{(\rm opt)}_m)^\top\mathbb J_{2N}\bm V^{(\rm opt)}_m
    =
    \mathbb J_{2m}.
\end{align}
Thus
\begin{align}
    \bm V^{(\rm opt)}_m
    \in
    \mathbb S(2m,2N).
\end{align}

Within the class \eqref{eq:ortho-symplectic-set}, $\bm V^{(\rm opt)}_m$ minimises the projection error \eqref{eq:proj-loss} \cite{buchfink2019symplecticmodelorderreduction}.  If the nonzero singular values of $\mathbb X_c$ are
\begin{align}
    \sigma_1\geq\sigma_2\geq\cdots\geq\sigma_r>0,
    \qquad
    r=\operatorname{rank}(\mathbb X_c),
\end{align}
then the optimal residual is
\begin{align}
    \min_{\bm V\in\mathbb S(2m,2N)}
    \big\|
        (\mathbf I-\bm V\bm V^+)\mathbb X_s
    \big\|_F
    =
    \left(
        \sum_{j=m+1}^{r}\sigma_j^2
    \right)^{1/2}.
    \label{eq:eym}
\end{align}
The singular values of the complex snapshot matrix therefore quantify the loss incurred by retaining only $m$ canonical pairs.  Large singular values identify symplectic directions strongly explored by the trajectory, while the discarded tail in \eqref{eq:eym} is precisely the part of the sampled motion that cannot be represented in the chosen $2m$-dimensional reduced phase space.  The cSVD basis thus turns the question of dynamical dimensionality into a controlled rank and singular-value problem.

\subsection{Reduced Hamiltonian dynamics}
\label{sec:reduced-hamiltonian-dynamics}
\label{sec:ROM}

The basis $\bm V$ does more than compress the sampled states.  Because it is symplectic, it also defines a Hamiltonian reduced-order model.  The reduced Hamiltonian is the pullback of the full Hamiltonian,
\begin{align}
    \widetilde{\mathcal H}(\widetilde{\bm z})
    :=
    \mathcal H(\bm V\widetilde{\bm z}).
    \label{eq:reduced-hamiltonian}
\end{align}
The reduced equations of motion are
\begin{align}
    \dot{\widetilde{\bm z}}
    =
    \mathbb J_{2m}\nabla_{\widetilde{\bm z}}
    \widetilde{\mathcal H}(\widetilde{\bm z})
    =
    \bm V^+X_{\mathcal H}(\bm V\widetilde{\bm z}),
    \qquad
    \widetilde{\bm z}(t_0)=\bm V^+\bm z_0.
    \label{eq:ROM}
\end{align}
The reconstructed full trajectory is
\begin{align}
    \bm z_{\mathrm{rec}}(t)
    =
    \bm V\widetilde{\bm z}(t).
    \label{eq:reconstructed-trajectory}
\end{align}

It is useful to distinguish two notions of exactness.  If the projection error vanishes on the snapshot matrix, then the sampled states themselves are represented exactly by the reduced subspace.  Exact reproduction by the autonomous reduced Hamiltonian system is stronger: the reduced subspace must also be invariant under the full Hamiltonian flow along the trajectory, or equivalently the full Hamiltonian vector field must be tangent to the embedded subspace.  For the quadratic free-field Hamiltonian studied below, this condition is satisfied once the observation window has resolved all normal-mode frequencies excited by the initial data.  The resulting ROM then reproduces the trajectory exactly, not merely at the sampled times.  In weakly nonlinear systems, by contrast, the subspace is generally only approximately invariant over a finite time window, and the ROM gives a controlled finite-time approximation.

\subsection{Minimal symplectic dimension}
\label{sec:rank=dim}

We now turn the preceding construction into a definition of dynamical dimension.  For a prescribed relative projection tolerance $\varepsilon$, define
\begin{align}
    d_{\min}(\varepsilon;\bm z_0)
    :=
    2\min
    \left\{
        m
        \ \middle|\
        \frac{
            \left(\sum_{j=m+1}^{r}\sigma_j^2\right)^{1/2}
        }{
            \left(\sum_{j=1}^{r}\sigma_j^2\right)^{1/2}
        }
        \leq
        \varepsilon
    \right\}.
    \label{eq:dmin-epsilon-definition}
\end{align}
The factor of two is important: $d_{\min}$ is the dimension of the reduced symplectic phase space, so the corresponding number of canonical pairs is $d_{\min}/2$.

In the exact case $\varepsilon=0$, Eq.~\eqref{eq:eym} gives
\begin{align}
    d_{\min}(0;\bm z_0)
    =
    2\,\operatorname{rank}(\mathbb X_c)
    =
    2r .
    \label{eq:dmin-rank}
\end{align}
Thus the exact minimal symplectic dimension is determined by the rank of the complex snapshot matrix.

The rank is meaningful only after the observation window is long enough to resolve the distinct temporal components present in the trajectory.  If the signal is quasi-periodic with angular frequencies separated by a minimum spacing $\Delta\Omega$, the finite-time frequency resolution gives the parametric requirement
\begin{align}
    t_1-t_0
    \gg
    \frac{\pi}{\Delta\Omega}.
    \label{eq:timewindow-resolution}
\end{align}
Once this resolution is reached, the rank of $\mathbb X_c$ stabilises.  For the UV/IR-regularised free scalar field, the temporal frequencies appearing in the trajectory are precisely the normal-mode frequencies of the field.  Therefore the exact minimal symplectic dimension reduces to a normal-frequency counting problem.

This is the main conceptual bridge used below. SMOR does not impose area scaling, or any dimensional reduction, by assumption.  It supplies a structure-preserving diagnostic: if a trajectory can be reproduced by a smaller autonomous Hamiltonian system, the singular values of $\mathbb X_c$ detect this, and $d_{\min}$ measures its symplectic dimension.  In the free scalar field, this diagnostic becomes analytically computable and leads to the frequency-counting results of the next section.

\section{Application to Scalar Field Theory}
\label{sec:application}

We now apply the SMOR framework developed in Sec.~\ref{sec:SMOR} to a concrete and physically motivated system: a classical real scalar field defined on a bounded spatial region, equipped with infrared (IR) and ultraviolet (UV) regulators. After regularisation, the phase space is finite-dimensional but large, with dimension $2N$ scaling proportionally to the regulated spatial volume.

Our guiding principle is deliberately dynamical.  For fixed physical parameters and a fixed initial condition, we regard as \emph{dynamically relevant} only those canonical directions that are actually explored by the Hamiltonian flow generated from that initial state.  Different initial conditions may explore different invariant sets, but in all cases the motion is confined to a lower-dimensional invariant subset of the full phase space \cite{marsden1999,Abraham1978FoundationsOM,arnold1989mathematical}.

The central question addressed in this section is the following: \emph{how large must a reduced Hamiltonian system be in order to reproduce the exact dynamics of a single scalar-field trajectory, when the reduction is required to preserve the standard symplectic structure?} Using the notion of minimal symplectic dimension introduced in Sec.~\ref{sec:SMOR}, this question admits a precise and operational answer.
\subsection{Summary of main findings}
\label{sec:summary}

Before presenting the detailed analysis, we summarise the main results obtained by applying the SMOR diagnostics of Sec.~\ref{sec:SMOR} to UV/IR--regularised scalar--field dynamics.

Operationally, for fixed physical parameters and a fixed initial condition, we sample the corresponding Hamiltonian trajectory over a finite observation window $[0,T_{\rm obs}]$ and construct the associated snapshot matrix. We then identify the smallest linear, time--independent symplectic subspace into which the trajectory can be projected with negligible error over this window. The dimension of this subspace defines the number of \emph{dynamically relevant} canonical degrees of freedom required to reproduce the motion by an autonomous reduced Hamiltonian system on the timescale probed. In all simulations reported here, the observation window is chosen to scale with the infrared length of the system, $T_{\rm obs}\sim\mathcal O(L_{\rm IR})$, which sets the natural longest dynamical timescale of the regulated field theory. In practice we take $T_{\rm obs}$ to be several times this scale in order to clearly resolve the asymptotic frequency content of the trajectory; the role of this choice and its implications for finite--time frequency resolution are discussed in detail in Secs.~\ref{sec:count-maxsym} and~\ref{sec:weak-interactions}.

Our first main result concerns the free scalar field. In this case, the dynamically relevant subspace is vastly smaller than the full kinematic phase space. Its dimension is not controlled by the number of discretised field variables, but instead by the number of \emph{distinct normal--mode frequencies} below the UV cutoff, each frequency counted once irrespective of degeneracy. This frequency--based mechanism is the structural reason why a drastic reduction of effective phase--space dimension is possible despite the volume--extensive size of the theory, and it underlies all scaling results reported below. For the free theory, the precise amplitudes and phases of the initial condition are immaterial for this counting: for generic initial data, the same minimal dimension is obtained.

\begin{figure}[t]
  \centering

  \begin{subfigure}{0.75\textwidth}
    \centering
    \includegraphics[width=\textwidth]{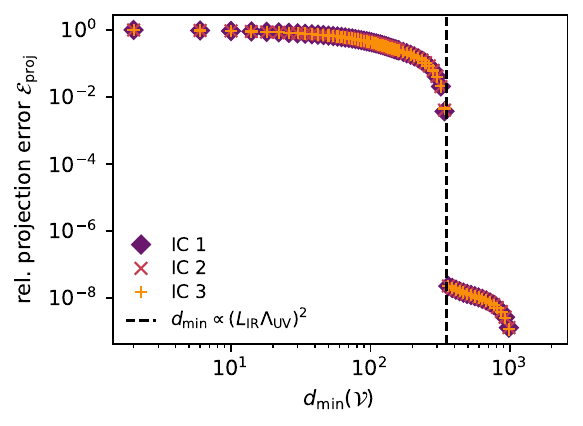}
    \caption{Free theory ($\lambda=0$).}
    \label{fig:projection_error_freefield}
  \end{subfigure}

  \medskip

  \begin{subfigure}{0.75\textwidth}
    \centering
    \includegraphics[width=\textwidth]{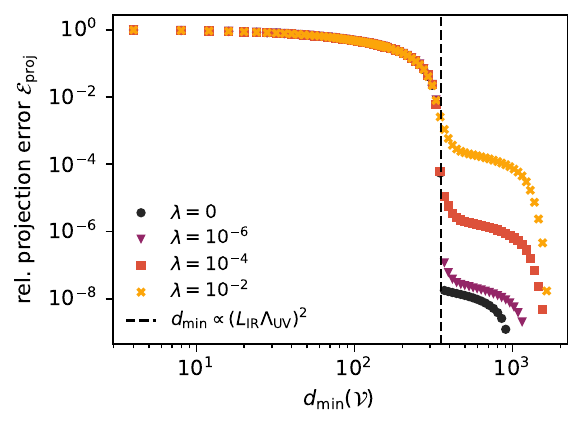}
    \caption{Weakly interacting $\lambda\phi^4$ theory.}
    \label{fig:projection_error_lambda}
  \end{subfigure}

  \caption{Relative projection error versus reduced dimension for the scalar field, extracted from trajectories sampled over a finite observation window $T_{\rm obs}$. 
(a) Free theory ($\lambda=0$): the error drops sharply at the threshold $d_{\min}=4\,n_\Omega \propto (L_{\rm IR}\Lambda_{\rm UV})^2$, independent of the (generic) Gaussian random initial condition. 
(b) Weakly interacting $\lambda\phi^4$ theory: initial amplitudes are fixed at the ground--state scale of the corresponding harmonic oscillators. The sharp jump is smoothed into a crossover due to finite--time frequency resolution, but the knee remains close to the free--field threshold for small $\lambda$.}
  \label{fig:projection_error}
\end{figure}

This structure manifests itself sharply in the projection error. As the reduced dimension is increased, the error drops abruptly at a well--defined threshold and then falls to numerical zero. This ``knee'' identifies the minimal symplectic dimension $d_{\min}$ required to represent the trajectory exactly on the observation window. In the free--field case, we find
\begin{align}
  d_{\min}=4\,n_\Omega,\nonumber
\end{align}
where $n_\Omega$ denotes the number of distinct normal--mode frequencies below the UV cutoff. Figure~\ref{fig:projection_error_freefield} illustrates this behaviour for several independent Gaussian random initial conditions in a flat spatial region. While the \emph{orientation} of the reduced subspace inside the full phase space depends on the initial condition, the threshold value itself is robust for generic data. The sharpness of the transition reflects the exact integrability of the free theory and the finite--time frequency resolution set by $T_{\rm obs}$.

This frequency counting leads directly to geometric scaling laws. In flat space, the number of distinct mode frequencies below a spherical UV cutoff grows proportionally to the boundary area of the region, up to slowly varying subleading corrections. As a consequence, the minimal symplectic dimension inherits an area--type scaling. When spatial curvature is introduced, the Laplacian spectrum is modified in a controlled manner: positively curved regions exhibit mild super--area growth, while negatively curved regions show sub--area behaviour, with the flat result recovered smoothly in the small--curvature limit.

To assess robustness beyond the exactly quadratic case, we also study weak self--interactions of the form $\lambda\phi^4$. In this case, the choice of initial amplitudes becomes physically relevant. We therefore initialise the interacting system with amplitudes set by the ground--state variances of the corresponding harmonic oscillators, ensuring low occupation numbers and weak nonlinear frequency shifts. As shown in Fig.~\ref{fig:projection_error_lambda}, the sharp error drop of the free theory is replaced by a smooth crossover, reflecting interaction--induced frequency shifts relative to the finite temporal resolution set by $T_{\rm obs}$. Nevertheless, the location of the knee remains anchored near the free--field threshold for small $\lambda$, so that the minimal symplectic dimension remains $d_{\min}\approx4\,n_\Omega$ on the timescales probed.

Finally, the reduced systems obtained from SMOR exhibit additional internal structure. The optimal reduced subspace decomposes into independent oscillator blocks, while linear combinations of these blocks define a larger set of ``apparent'' field modes. These apparent modes are not mutually independent: their Poisson brackets are governed by a projector rather than the identity, reflecting the fact that they share the same reduced canonical degrees of freedom. Upon canonical quantisation, this structure would translate into nontrivial overlap patterns among the corresponding operators. A closely related overlap structure has recently appeared in quantum field--theoretic constructions \cite{Friedrich2024}, but here it arises directly from classical symplectic reduction rather than from modified quantum commutation relations.

The remainder of this section derives these results in detail. We first analyse the free scalar field in flat and curved spatial regions, showing analytically how the Laplacian spectrum controls the minimal symplectic dimension. We then examine the stability of this picture under weak interactions and discuss the emergent overlap structure of the reduced degrees of freedom.

\subsection{Regularised free scalar field as a finite Hamiltonian system}
\label{sec:setup}

We now specify the finite-dimensional Hamiltonian systems to which the SMOR construction of Sec.~\ref{sec:SMOR} will be applied.  The starting point is a real free scalar field on a static product spacetime $\mathbb R\times\Sigma$, restricted to a bounded spatial region and truncated at a finite ultraviolet scale.  The infrared regulator is the finite region, while the ultraviolet regulator is imposed by retaining only finitely many eigenmodes of the spatial Laplacian.  After these two regulators are imposed, the field theory becomes an ordinary finite-dimensional Hamiltonian system with phase space $\mathcal M\simeq\mathbb R^{2N}$.

The only structural fact we need is the standard one: in an orthonormal basis of spatial Laplacian eigenfunctions, the free scalar Hamiltonian becomes a sum of decoupled harmonic oscillators.  The spatial geometry and the boundary condition enter only through the Laplacian eigenvalues, and hence through the oscillator frequencies.  This is why the reduction problem below becomes a question about the distinct normal-mode frequencies present below the UV cutoff.

We consider two representative settings. First, we use a periodic cubic box in flat space, where the spectrum is explicit and leads to the arithmetic counting problem analysed in Sec.~\ref{sec:count-flat}. Second, we consider geodesic balls in three-dimensional maximally symmetric spaces, which allow us to isolate how spatial curvature changes the frequency spectrum.  Throughout this part of the paper the spatial geometry is time independent; extending the analysis to FRW backgrounds, where the mode frequencies become explicitly time dependent, is left for future work.

\subsubsection{Flat box regulator in  Minkowski space}
\label{sec:setup-box}

Let $\phi(t,\bm x)$ be a real scalar field in a cubic box
\begin{align}
    \mathcal B
    =
    [-L_{\mathrm{IR}}/2,L_{\mathrm{IR}}/2]^3
\end{align}
with periodic boundary conditions.  The Fourier momenta are
\begin{align}
    \bm k
    =
    \frac{2\pi}{L_{\mathrm{IR}}}\bm n,
    \qquad
    \bm n\in\mathbb Z^3,
\end{align}
and we impose the spherical UV cutoff
\begin{align}
    |\bm k|\leq \Lambda_{\mathrm{UV}}.
\end{align}
Equivalently, we retain only finitely many Fourier modes of the field.  Expanding
\begin{align}
    \phi(t,\bm x)
    =
    \sum_{|\bm k|\leq\Lambda_{\mathrm{UV}}}
    \phi_{\bm k}(t)e^{i\bm k\cdot\bm x},
    \qquad
    \phi_{-\bm k}=\phi_{\bm k}^{\ast},
\end{align}
and using the standard free scalar Lagrangian, the truncated theory is a finite collection of harmonic modes with frequencies
\begin{align}
    \Omega_{\bm k}
    =
    \sqrt{|\bm k|^2+m_0^2}.
\end{align}

Since the field is real, the complex Fourier coefficients are not independent.  Choosing any equivalent set of independent real canonical coordinates $(Q_{\bm k},P_{\bm k})$, defined by
\begin{align}
  \Omega_{\bm k} Q_{\bm k}
  &:=
  \Omega_{\bm k}\,\mathrm{Re}(\phi_{\bm k})
  -
  \mathrm{Im}(\pi_{\bm k}),\nonumber
  \\
  P_{\bm k}
  &:=
  \Omega_{\bm k}\,\mathrm{Im}(\phi_{\bm k})
  +
  \mathrm{Re}(\pi_{\bm k}),
  \label{eq:PQ_def}
\end{align}
with inverse relation
\begin{align}
  \phi_{\bm k}
  =
  \frac{1}{2\Omega_{\bm k}}
  \Big[
  \Omega_{\bm k}(Q_{\bm k}+Q_{-\bm k})
  +
  i\,(P_{\bm k}-P_{-\bm k})
  \Big]\ ,
\end{align}
the Hamiltonian can be written in the oscillator form
\begin{align}
    \mathcal H
    =
    \frac12
    \sum_{|\bm k|\leq\Lambda_{\mathrm{UV}}}
    \left(
        P_{\bm k}^2
        +
        \Omega_{\bm k}^2 Q_{\bm k}^2
    \right),
    \label{eq:H_box}
\end{align}
where the sum is understood as a sum over independent real modes.  Thus the regulated free field is a finite-dimensional quadratic Hamiltonian system.  Its full phase-space dimension is $2N$, where $N$ is the number of retained canonical oscillator modes.

For a spherical cutoff in Fourier space, the usual mode count grows with the volume of the allowed momentum-space ball:
\begin{align}
    N(L_{\mathrm{IR}},\Lambda_{\mathrm{UV}})
    \approx
    \frac{\tfrac{4}{3}\pi\Lambda_{\mathrm{UV}}^3}
         {(2\pi/L_{\mathrm{IR}})^3}
    =
    \frac{(L_{\mathrm{IR}}\Lambda_{\mathrm{UV}})^3}{6\pi^2}.
    \label{eq:volume-scaling}
\end{align}

This is the ordinary kinematical regulator count. Equivalently, it is the leading Weyl-law count of Laplacian eigenmodes below the cutoff \cite{Ivrii_2016}, where eigenvalues are counted with their multiplicities. The central question of the paper is different. We do not ask how many mode labels are retained by the regulator, but how many distinct canonical oscillator directions are dynamically needed to reproduce a trajectory. For the free theory, this number is controlled by the distinct frequencies $\Omega_{\bm k}$ excited below the cutoff, with degeneracies stripped away.

\subsubsection{Geodesic balls in maximally symmetric spaces}
\label{sec:setup-maxsym}

We also consider the same free scalar field on a geodesic ball
\begin{align}
    \mathcal B
    =
    \{\rho\leq \rho_{\mathrm{IR}}\}
\end{align}
cut from a three-dimensional maximally symmetric space of constant sectional curvature
\begin{align}
    K\in\{0,\pm 1/R_c^2\}.
\end{align}
In geodesic polar coordinates, the spatial metric is
\begin{align}
    \gamma_{ij}\dd x^i\dd x^j
    =
    \dd\rho^2
    +
    S_K^2(\rho)
    \left(
        \dd\theta^2+\sin^2\theta\,\dd\varphi^2
    \right),
\end{align}
with
\begin{align}
    S_K(\rho)
    =
    \begin{cases}
        R_c\sin(\rho/R_c), & K>0,\\
        \rho, & K=0,\\
        R_c\sinh(\rho/R_c), & K<0.
    \end{cases}
\end{align}
At $\rho=\rho_{\mathrm{IR}}$ we impose a choice of Dirichlet, Neumann, or Robin boundary condition.

Let $-\Delta_\gamma$ be the Laplace--Beltrami operator on the ball.  We choose an orthonormal eigenbasis
\begin{align}
    -\Delta_\gamma u_\alpha
    =
    k_\alpha^2 u_\alpha,
    \qquad
    \int_{\mathcal B}\sqrt{\gamma}\,
    u_\alpha^\ast u_\beta\,\dd^3x
    =
    \delta_{\alpha\beta},
    \label{eq:LB-orthonormal}
\end{align}
and retain only modes with
\begin{align}
    k_\alpha\leq \Lambda_{\mathrm{UV}}.
\end{align}
For geodesic balls in maximally symmetric spaces, separation of variables gives
\begin{align}
    \alpha=(n,\ell,m),
    \qquad
    u_{n\ell m}(\bm x)
    =
    R_{n\ell}(\rho)Y_{\ell m}(\theta,\varphi),
\end{align}
with $m=-\ell,\ldots,\ell$.  The radial functions satisfy
\begin{align}
    \frac{1}{S_K^2(\rho)}
    \frac{\dd}{\dd\rho}
    \left(
        S_K^2(\rho)
        \frac{\dd R_{n\ell}}{\dd\rho}
    \right)
    -
    \frac{\ell(\ell+1)}{S_K^2(\rho)}
    R_{n\ell}
    +
    k_{n\ell}^2 R_{n\ell}
    =
    0,
    \label{eq:radialeq}
\end{align}
together with the chosen boundary condition at $\rho=\rho_{\mathrm{IR}}$.

Expanding the scalar field in this eigenbasis,
\begin{align}
    \phi(t,\bm x)
    =
    \sum_{k_\alpha\leq\Lambda_{\mathrm{UV}}}
    \phi_\alpha(t)u_\alpha(\bm x),
\end{align}
the free scalar Hamiltonian again reduces to a sum of decoupled oscillators.  In real canonical variables $(Q_\alpha,P_\alpha)$, defined in direct
analogy with the flat–space case, one obtains
\begin{align}
    \mathcal H
    =
    \frac12
    \sum_{k_\alpha\leq\Lambda_{\mathrm{UV}}}
    \left(
        P_\alpha^2
        +
        \Omega_\alpha^2 Q_\alpha^2
    \right),
    \qquad
    \Omega_\alpha
    =
    \sqrt{k_\alpha^2+m_0^2}.
    \label{eq:H_general_modes}
\end{align}
Thus the Hamiltonian form is universal: all dependence on curvature, boundary conditions, and the shape of the region is encoded in the spectrum $\{k_\alpha\}$. The usual Weyl count of retained eigenfunctions is again a multiplicity-counted spectral count and is volume-extensive at leading order \cite{Ivrii_2016}. The SMOR rank of a free trajectory, however, is governed by the distinct oscillator frequencies $\{\Omega_\alpha\}$ present below the UV cutoff. The following sections use this distinction to turn dynamical dimension counting into a degeneracy-stripped spectral counting problem.

\subsection{Minimal reduced symplectic dimension for a free scalar field}
\label{sec:MinSymplDim}

We now determine the minimal symplectic dimension required to reproduce the Hamiltonian trajectory generated by a \emph{single} initial condition of the regularised free scalar field. This quantity is precisely what the SMOR construction detects in the free theory and will serve as the reference point for the subsequent analysis of geometry and weak interactions.

For a quadratic Hamiltonian the underlying mechanism is particularly transparent. Because the equations of motion are linear, the time evolution is quasi-periodic and remains confined to a finite-dimensional invariant symplectic subspace of the full phase space. Crucially, the dimension of this subspace is not controlled by the total number of retained modes, which scales with volume, but by the set of \emph{distinct physical frequencies} present below the UV cutoff. This structural property is the reason why a drastic reduction of the effective phase-space dimension is possible despite the kinematically large size of the regularised field theory.

In mode coordinates the free Hamiltonian takes the quadratic form
\begin{align}
  \mathcal H(\bm Q,\bm P)  =  \frac12\,\bm P^\top \bm P  +  \frac12\,\bm Q^\top \Omega^2 \bm Q,  \qquad  \Omega^2=\mathrm{diag}(\Omega_\alpha^2),
\end{align}
where $\alpha$ labels the retained normal modes. Introducing the phase-space vector $\bm z:=(\bm Q,\bm P)\in\mathbb R^{2N}$, Hamilton’s equations can be written as the linear system
\begin{align}
  \dot{\bm z}(t) = \mathbb A\,\bm z(t),  \qquad  \mathbb A  =  \begin{pmatrix}    0 & \mathbb I_N \\
    -\Omega^2 & 0
  \end{pmatrix},
\end{align}
or equivalently $\mathbb A=\mathbb J_{2N}\nabla^2\mathcal H$, with $\mathbb J_{2N}$ the canonical symplectic matrix. The solution is therefore
\begin{align}
  \bm z(t)=\exp(t\mathbb A)\,\bm z_0 .
\end{align}
Because the flow is linear and generated by a time-independent matrix, the smallest linear subspace containing the trajectory of a given initial condition is automatically invariant under the full time evolution. Once such a subspace is identified from snapshots on a sufficiently long time window, it therefore contains the trajectory for all times, not merely for the sampled interval. This is the sense in which, for quadratic Hamiltonians, SMOR can become \emph{exact}: the reduced space is an invariant subspace selected by the dynamics, rather than a best-fit approximation tied to a finite observation window.

Each normal mode labelled by $\alpha$ evolves independently according to
\begin{align}
  \dot Q_\alpha &= P_\alpha,
  \qquad
  \dot P_\alpha = -\,\Omega_\alpha^2\, Q_\alpha ,
\end{align}
with general solution
\begin{align}
  Q_\alpha(t)
  &= A_\alpha \cos(\Omega_\alpha t)
     + B_\alpha \sin(\Omega_\alpha t), \nonumber\\
  P_\alpha(t)
  &= -\,\Omega_\alpha A_\alpha \sin(\Omega_\alpha t)
     + \Omega_\alpha B_\alpha \cos(\Omega_\alpha t),
\end{align}
where the real coefficients $A_\alpha$ and $B_\alpha$ are fixed by the initial condition.

Distinct mode labels $\alpha$ may share the same physical frequency. It is therefore convenient to group modes into sets of equal frequency, which we refer to as frequency shells. Let $S$ denote the index set of distinct frequencies present below the UV cutoff, and for each $s\in S$ let $\Omega_s$ be the corresponding frequency. We then define the associated shell by
\begin{align}
  \mathcal K_s := \{\alpha \mid \Omega_\alpha = \Omega_s\},
  \qquad
  g_s := |\mathcal K_s|.
\end{align}

Collecting the canonical variables belonging to a fixed shell $s$ into vectors
\begin{align}
  Q_s(t)\in\mathbb R^{g_s},
  \qquad
  P_s(t)\in\mathbb R^{g_s},
\end{align}
the Hamiltonian restricted to that shell takes the form
\begin{align}
  H_s
  \;=\;
  \frac12\big(\|P_s\|^2 + \Omega_s^2 \|Q_s\|^2\big),
\end{align}
which is isotropic in the internal space $\mathbb R^{g_s}$.

The solution within a given shell can be written compactly as
\begin{align}
  Q_s(t)
  &=
  A_s \cos(\Omega_s t)
  +
  B_s \sin(\Omega_s t), \nonumber\\
  P_s(t)
  &=
  -\,\Omega_s A_s \sin(\Omega_s t)
  +
  \Omega_s B_s \cos(\Omega_s t),
\end{align}
with fixed vectors $A_s,B_s\in\mathbb R^{g_s}$ determined by the initial data. For each distinct frequency $\Omega_s$, the motion is therefore confined to the two--dimensional subspace
\begin{align}
  \mathcal U_s := \mathrm{span}\{A_s,B_s\}
  \subset \mathbb R^{g_s}.
\end{align}

Independently of the degeneracy $g_s$, the trajectory explores only two configuration--space directions within each shell, together with their two conjugate momenta. Equivalently, each distinct frequency $\Omega_s$ contributes a four--dimensional invariant symplectic block to the phase space. The sine and cosine functions merely provide a convenient basis for this motion; they correspond to the two independent time dependences required to represent a single--frequency orbit.

This structure is mirrored exactly in the snapshot construction. Let
\begin{align}
  \mathbb X_c
  :=
  \mathbb X_s^{(Q)} + i\,\mathbb X_s^{(P)} \in \mathbb C^{N\times M}
\end{align}
denote the complex snapshot matrix built from samples at times $t_1<\dots<t_M$. For a fixed frequency $\Omega$, introduce the sampling vectors
\begin{align}
  \bm s_\Omega
  :=
  \big(\sin(\Omega t_1),\dots,\sin(\Omega t_M)\big)^\top,
  \qquad
  \bm c_\Omega
  :=
  \big(\cos(\Omega t_1),\dots,\cos(\Omega t_M)\big)^\top .
\end{align}
For any $\alpha\in\mathcal K_\Omega$, the corresponding row of $\mathbb X_c$ is a linear combination of $\bm s_\Omega^\top$ and $\bm c_\Omega^\top$. Hence all rows belonging to a given shell lie in a two-dimensional subspace of $\mathbb C^M$. Summing over shells, the rank of the complex snapshot matrix is therefore bounded by
\begin{align}
  \mathrm{rank}(\mathbb X_c) \le 2\,n_\Omega,
\end{align}
where $n_\Omega$ is the number of distinct frequencies below the UV cutoff. For generic sampling times and generic initial conditions, both sine and cosine components are present for each excited frequency, and the vectors $\bm s_\Omega$ and $\bm c_\Omega$ are linearly independent. In this sense the bound is maximally saturated,
\begin{align}
  \mathrm{rank}(\mathbb X_c)=2\,n_\Omega,\nonumber
\end{align}
and no extension of the sampling window or change of initial state can increase this rank within the quadratic theory, since the Hamiltonian flow generates no additional independent time dependences.

As reviewed in Sec.~\ref{sec:SMOR}, choosing $m=\mathrm{rank}(\mathbb X_c)$ in the complex-SVD construction yields an ortho--symplectic embedding of real dimension $2m$. Combining this with the saturated rank immediately gives the minimal symplectic dimension required to reproduce the trajectory exactly,
\begin{align}
  d_{\min}=2m=4\,n_\Omega.
\end{align}
The remaining task is therefore to determine $n_\Omega$, the number of distinct frequencies below the UV cutoff, for the regulators introduced in Sec.~\ref{sec:setup}. This is carried out first for a flat periodic box in Sec.~\ref{sec:count-flat}, and then for geodesic balls in maximally symmetric curved spaces in Sec.~\ref{sec:count-maxsym}.

\subsection{Counting distinct frequencies in a flat box}
\label{sec:count-flat}

We now derive an explicit expression for the number of distinct physical frequencies $n_\Omega$ of the regularised free scalar field in a flat spatial box. The usual Weyl-law count would count all Fourier modes below the UV cutoff, including degeneracies, and hence gives the volume-extensive estimate \eqref{eq:volume-scaling}. Here we instead count the distinct values of the frequency. As shown below, this degeneracy-stripped count reduces to the classical three-square problem, allowing us to invoke Legendre's three-square theorem \cite{Ankeny1957ThreeSquares,Pollack2018DirichletsPO}. This provides analytic control over the minimal symplectic dimension detected by SMOR and explains why it scales with the \emph{area} of the IR/UV window, up to subleading corrections.

\medskip

We work with the periodic cubic box of side length $L_{\mathrm{IR}}$ introduced in Sec.~\ref{sec:setup-box}. Fourier modes are labelled by integer triples $\bm n\in\mathbb Z^3$,
\begin{align}
  k_i &= \frac{2\pi}{L_{\mathrm{IR}}}\,n_i,
  \qquad n_i\in\mathbb Z,
  \\
  \Omega_{\bm k}^2 &= |\bm k|^2 + m_0^2
  = k_1^2+k_2^2+k_3^2 + m_0^2 .
\end{align}
Imposing a spherical UV cutoff $|\bm k|\le\Lambda_{\mathrm{UV}}$ is equivalent to the integer constraint
\begin{align}
  n_1^2+n_2^2+n_3^2 \;\le\; X,
  \qquad
  X := \Big(\frac{L_{\mathrm{IR}}\Lambda_{\mathrm{UV}}}{2\pi}\Big)^2 .
\end{align}

Distinct physical frequencies correspond to distinct values of $|\bm k|^2$, or equivalently to distinct integers of the form
\begin{align}
  n = n_1^2+n_2^2+n_3^2 \in \{0,1,\dots,X\}.
\end{align}
We therefore introduce the set of realised squared wave numbers
\begin{align}
  \mathcal{R}(X)
  :=
  \big\{
    n\in\{0,1,\dots,X\}
    \;:\;
    n=x^2+y^2+z^2 \text{ for some } x,y,z\in\mathbb Z
  \big\},
\end{align}
so that the number of distinct frequencies below the cutoff is
\begin{align}
  n_\Omega = |\mathcal{R}(X)|.
\end{align}
The set $\mathcal{R}(X)$ is characterised by Legendre’s three–square theorem \cite{Ankeny1957ThreeSquares,Pollack2018DirichletsPO}, which states that a non-negative integer $n$ can be written as a sum of three squares if and only if it is \emph{not} of the form
\begin{align}
  n = 4^a(8b+7),
  \qquad a,b\in\mathbb N_0 .
\end{align}
Defining the excluded set
\begin{align}
  L(X)
  :=
  \big\{
    n\le X \;:\; n=4^a(8b+7)
    \text{ for some } a,b\ge 0
  \big\},
\end{align}
one obtains the exact identity
\begin{align}
  |\mathcal{R}(X)| = X+1 - |L(X)|.
\end{align}

To estimate $|L(X)|$, we decompose it as a disjoint union $L(X)=\bigsqcup_{a=0}^{a_{\max}} A_a$ with
\begin{align}
  A_a
  &:= \big\{\,4^a(8b+7)\le X:\ b\in\mathbb Z_{\ge 0}\,\big\},\nonumber
  \\
  a_{\max}
  &= \left\lfloor \log_4\!\left(\frac{X}{7}\right)\right\rfloor .
\end{align}
Each subset $A_a$ contains
\begin{align}
  |A_a|
  &= \left\lfloor \frac{X-7\cdot 4^a}{8\cdot 4^a}\right\rfloor
  = \frac{X}{8\cdot 4^a} + \mathcal O(1)
\end{align}
elements. Summing over $a$ yields
\begin{align}
  |L(X)|
  &= \sum_{a=0}^{a_{\max}} |A_a|\nonumber
  \\
  &=
  \frac{X}{8}
  \sum_{a=0}^{a_{\max}}\left(\frac{1}{4}\right)^a
  + \mathcal O(a_{\max})\nonumber
  \\
  &\simeq
  \frac{X}{6}
  - \eta \log_4\!\left(\frac{X}{7}\right),
  \qquad
  \eta\sim\mathcal O(1),
\end{align}
for large $X$. Consequently,
\begin{align}
  |\mathcal{R}(X)|
  \;\simeq\;
  \frac{5}{6}\,X
  + \eta \log_4\!\left(\frac{X}{7}\right),
  \qquad
  X\to\infty .
  \label{eq:R-asymptotic}
\end{align}

\begin{figure}[t]
  \centering
  \includegraphics[width=0.8\textwidth]{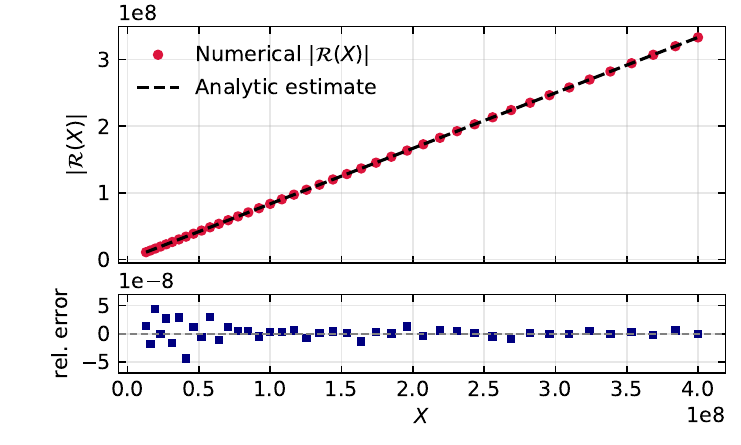}
  \caption{
    Verification of the three--square counting estimate.
    \emph{Top:} numerical count $|\mathcal{R}(X)|$ of integers
    $n\le X$ representable as a sum of three squares (red points),
    compared with the analytic asymptotic~\eqref{eq:R-asymptotic}
    (black dashed line).
    \emph{Bottom:} relative error between the numerical count and the
    analytic estimate, which remains at the level of $\mathcal{O}(10^{-8})$
    over the range of $X$ relevant for our simulations.
  }
  \label{fig:legendre_verification}
\end{figure}

Figure~\ref{fig:legendre_verification} shows a numerical verification of this asymptotic estimate over the range of $X$ relevant for our simulations.

Finally, recalling that $X=(L_{\mathrm{IR}}\Lambda_{\mathrm{UV}}/2\pi)^2$ and that the minimal symplectic dimension is $d_{\min}=4\,n_\Omega$, we obtain
\begin{align}
    d_{\rm min}(\mathcal{V}) \simeq \frac{5}{6\pi^2}L_{\rm IR}^2\Lambda_{\rm UV}^2 + \tilde{\eta}\cdot\log_2\!\bigg[\frac{L_{\rm IR}^2\Lambda_{\rm UV}^2}{28\pi^2}\bigg]\ , \quad \tilde{\eta} \sim\mathcal{O}(1)\ .
\end{align}
Equivalently, using $|\partial\mathcal B|=6L_{\mathrm{IR}}^2$ for the boundary area of the cube,
\begin{align}
  d_{\min}
  &\simeq
  \frac{5}{36\pi^2}\,|\partial\mathcal B|\,\Lambda_{\mathrm{UV}}^2
  + \mathcal O\!\big(\log(L_{\mathrm{IR}}\Lambda_{\mathrm{UV}})\big).
\end{align}
Thus, while the kinematic phase–space dimension grows volume-extensively, $\dim\mathcal M\sim L_{\mathrm{IR}}^3\Lambda_{\mathrm{UV}}^3$, the number of dynamically relevant canonical directions required to reproduce a single trajectory grows only with the boundary area (up to subleading logarithmic corrections).

\subsection{Counting distinct frequencies in maximally symmetric spaces}
\label{sec:count-maxsym}

Having established in Sec.~\ref{sec:count-flat} how the number of distinct mode frequencies scales in a flat periodic box, we now repeat the analysis for curved spatial geometries.  Throughout this subsection the field content and UV/IR regulators are kept fixed; only the \emph{spatial geometry} is varied.  Our goal is to determine how curvature modifies the counting of distinct frequencies $n_\Omega$, and hence the minimal symplectic dimension $d_{\min}=4\,n_\Omega$ detected by SMOR.

As a useful consistency check, we will first consider the Euclidean ball ($K=0$). This differs from Sec.~\ref{sec:count-flat} only by the IR regulator (ball rather than periodic box), while sharing the same local flat geometry.  We then turn to the genuinely curved cases $K\neq 0$, where curvature induces controlled sub-/super-area deviations.

We consider a geodesic ball $\mathcal B=\{\rho\le\rho_{\mathrm{IR}}\}$ cut from a three–dimensional maximally symmetric space of constant sectional curvature $K\in\{0,\pm 1/R_c^2\}$. As shown in Sec.~\ref{sec:setup-maxsym}, the scalar field decomposes into normal modes labelled by $(n,\ell,m)$ with frequencies $\Omega_{n\ell}=\sqrt{k_{n\ell}^2+m_0^2}$, where the spatial wavenumbers $k_{n\ell}$ are determined by the radial eigenvalue problem \eqref{eq:radialeq}.  The number of \emph{distinct} frequencies below the UV cutoff therefore coincides with the number of distinct radial eigenvalues $k_{n\ell}\le\Lambda_{\mathrm{UV}}$, ignoring the $m$–degeneracy.

To estimate this number we follow Refs.~\cite{kosowsky1998efficientcomputationhypersphericalbessel,tram2017computationhypersphericalbesselfunctions} and use a WKB approximation for the radial equation. Introducing the radial “momentum” $p_\rho(\rho;k,\ell)$, defined by rewriting \eqref{eq:radialeq} in Schrödinger–like form, the leading WKB quantisation condition reads
\begin{align}
  \int_{\rho_\ast}^{\rho_{\mathrm{IR}}}
  p_\rho(\rho;k_{n\ell},\ell)\,\dd\rho
  \;=\;
  \pi\big(n(\ell)+\alpha_{\mathrm{BC}}\big),
  \label{eq:wkb-general}
\end{align}
with
\begin{align}
  p_\rho^2(\rho;k,\ell)
  \;=\;
  k^2
  - \frac{(\ell+\tfrac12)^2}{S_K^2(\rho)}
  - K .
\end{align}
Here $\rho_\ast$ is the turning point defined by $p_\rho(\rho_\ast)=0$, $\alpha_{\mathrm{BC}}=\mathcal O(1)$ is a boundary (Maslov) phase depending on the choice of boundary condition \cite{Bender:1999box}, and the Langer correction $\ell(\ell+1)\to(\ell+\tfrac12)^2$ has been included.  We focus on the large–cutoff regime $\Lambda_{\mathrm{UV}}R_c\gg1$, where the leading scaling is controlled by the action integral.

\paragraph{Flat ball limit ($K=0$).}
For vanishing curvature the radial momentum reduces to
\begin{align}
  p_\rho^2(\rho;k,\ell)
  \;=\;
  k^2-\frac{(\ell+\tfrac12)^2}{\rho^2},
\end{align}
with turning point $\rho_\ast=(\ell+\tfrac12)/k$. For fixed $\ell$, the number of radial levels with $k_{n\ell}\le\Lambda_{\mathrm{UV}}=:k_{\max}$ is
\begin{align}
  n(\ell)
  \;\simeq\;
  \frac{1}{\pi}
  \int_{\rho_\ast}^{\rho_{\mathrm{IR}}}
  \sqrt{k_{\max}^2-\frac{(\ell+\tfrac12)^2}{\rho^2}}\;\dd\rho
  \;+\;\mathcal O(1).
\end{align}
Keeping only the leading contribution for $k_{\max}\rho_{\mathrm{IR}}\gg1$, this integral evaluates to
\begin{align}
  n(\ell)
  \;\simeq\;
  \frac{1}{\pi}\Big[
    k_{\max}\rho_{\mathrm{IR}}\sqrt{1-x^2}
    -(\ell+\tfrac12)\arcsin x
  \Big]_{x=\frac{\ell+\frac12}{k_{\max}\rho_{\mathrm{IR}}}}
  \;+\;\mathcal O(1).
\end{align}
Summing over allowed angular momenta and ignoring the $m$–degeneracy,
\begin{align}
  n_\Omega
  \;\simeq\;
  \sum_{\ell=0}^{\ell_{\max}} n(\ell)
  \;\approx\;
  \int_0^{k_{\max}\rho_{\mathrm{IR}}} n(\ell)\,\dd\ell .
\end{align}
Using
\begin{align}
  \int_0^1\!\sqrt{1-x^2}\,\dd x=\frac{\pi}{4},
  \qquad
  \int_0^1\!x\arcsin x\,\dd x=\frac{\pi}{8},
\end{align}
one finds the leading asymptotics
\begin{align}
  n_\Omega
  \;\simeq\;
  \frac{(k_{\max}\rho_{\mathrm{IR}})^2}{8}
  \;+\;\mathcal O(k_{\max}\rho_{\mathrm{IR}})
  \;=\;
  \frac{(\Lambda_{\mathrm{UV}}\rho_{\mathrm{IR}})^2}{8}
  \;+\;\mathcal O(\Lambda_{\mathrm{UV}}\rho_{\mathrm{IR}}).
\end{align}
Since the boundary area is $|\partial\mathcal B|=4\pi\rho_{\mathrm{IR}}^2$ and $d_{\min}=4\,n_\Omega$, this reproduces an \emph{area scaling} of the minimal symplectic dimension,
\begin{align}
  d_{\min}
  \;\simeq\;
  \frac{1}{2\pi}\,\frac{|\partial\mathcal B|\,\Lambda_{\mathrm{UV}}^2}{4}
  \;+\;\mathcal O(\Lambda_{\mathrm{UV}}\rho_{\mathrm{IR}}).
\end{align}
Boundary conditions affect only the subleading terms through $\alpha_{\mathrm{BC}}$ and do not modify the leading scaling.

\paragraph{Nonzero curvature ($K\neq0$).}
For constant curvature the function $S_K(\rho)$ modifies the radial momentum to
\begin{align}
  p_\rho^2(\rho;k,\ell;K)
  \;=\;
  k^2-\frac{\ell(\ell+1)}{S_K^2(\rho)}-K .
\end{align}
Carrying through the WKB analysis in the regime $k_{\max}R_c\gg1$ and retaining only the leading contribution, one finds for the minimal symplectic dimension
\begin{align}
  \text{Open }(K<0):\quad
  d_{\min}
  &\;\simeq\;
  \frac{(\Lambda_{\mathrm{UV}}R_c)^2}{2}
  \Bigg[4\sinh^2\!\Big(\frac{\rho_{\mathrm{IR}}}{2R_c}\Big)\Bigg],
  \label{eq:nOmega-open}\\[6pt]
  \text{Closed }(K>0):\quad
  d_{\min}
  &\;\simeq\;
  \frac{(\Lambda_{\mathrm{UV}}R_c)^2}{2}
  \Bigg[4\sin^2\!\Big(\frac{\rho_{\mathrm{IR}}}{2R_c}\Big)\Bigg].
  \label{eq:nOmega-closed}
\end{align}
Comparing with the corresponding boundary areas,
\begin{align}
  |\partial\mathcal B|_{K<0}
  &=4\pi R_c^2\sinh^2\!\Big(\frac{\rho_{\mathrm{IR}}}{R_c}\Big),\\
  |\partial\mathcal B|_{K>0}
  &=4\pi R_c^2\sin^2\!\Big(\frac{\rho_{\mathrm{IR}}}{R_c}\Big),
\end{align}
one obtains
\begin{align}
  \frac{d_{\min}}{\Lambda_{\mathrm{UV}}^2|\partial\mathcal B|}\Bigg|_{K<0}
  &\;\simeq\;
  \frac{1}{8\pi\cosh^2(\rho_{\mathrm{IR}}/2R_c)}
  \;\le\;\frac{1}{8\pi},
  \qquad\text{(sub–area suppression)},\\[4pt]
  \frac{d_{\min}}{\Lambda_{\mathrm{UV}}^2|\partial\mathcal B|}\Bigg|_{K>0}
  &\;\simeq\;
  \frac{1}{8\pi\cos^2(\rho_{\mathrm{IR}}/2R_c)}
  \;\ge\;\frac{1}{8\pi},
  \qquad\text{(super–area enhancement)}.
\end{align}
In the small–radius limit $\rho_{\mathrm{IR}}/R_c\to0$ both expressions reduce to the flat–ball result, as expected.  Curvature therefore deforms—but does not eliminate—the frequency–counting mechanism underlying SMOR, leading to geometry–dependent deviations from strict area scaling.

Let us finally discuss how the sub-area suppression and super-area enhancement found above should be interpreted in the presence of curvature. The results above highlight two general features of the SMOR reduction for classical field theories.

First, the observed area--type scaling of the minimal symplectic dimension $d_{\min}=4\,n_\Omega$ is a purely \emph{classical} phenomenon. It does not rely on entanglement, entropy, or gravitational backreaction. Rather, it reflects the fact that, for a given initial condition, the Hamiltonian dynamics explores only a low--rank set of distinct normal--mode frequencies compared to the volume--extensive number of canonical field variables. The SMOR construction makes this reduction precise by identifying the invariant symplectic subspace actually used by the trajectory.

Second, the curvature dependence of the scaling demonstrates that this low--rank structure is not arbitrary, but inherited directly from the spectral properties of the spatial Laplacian \cite{Chavel1984EigenvaluesIR}. In maximally symmetric spaces, positive curvature places more distinct radial eigenvalues below a fixed UV cutoff at large radius, leading to a mild super--area growth of $n_\Omega$, while negative curvature suppresses this growth and yields sub--area behaviour. In this sense, the number of dynamically relevant canonical directions tracks how the Laplacian spectrum is distributed in frequency space for a given geometry and boundary condition.

Finally, we would like to emphasise that the frequency counting discussed above refers to the exact spectrum of the UV--regulated free theory, for which the minimal symplectic dimension $d_{\min}=4\,n_\Omega$ follows sharply from the integrable dynamics. In practice, however, both numerical implementations and physical dynamical probes access trajectories only over a finite observation window $T_{\rm obs}$.  This implies a finite temporal frequency resolution
\begin{align}
  \Delta\Omega \;\sim\; \frac{\pi}{T_{\rm obs}},
\end{align}
so that free--theory frequencies whose separations lie below $\Delta\Omega$ cannot be operationally distinguished on that timescale.

In the box simulations shown in Fig.~\ref{fig:projection_error}, we evolve the free system for $T_{\rm obs}\simeq 14\,L_{\rm IR}$ in order to clearly expose the asymptotic behaviour of the SMOR projection error and the location of the characteristic ``knee'' associated with area--type scaling. The free--theory rank saturation at $d_{\min}=4\,n_\Omega$ remains sharply visible, up to a small rounding controlled by the finite resolution $\Delta\Omega$. This finite--time effect is not a limitation of the SMOR procedure but reflects the generic fact that exact frequency degeneracies are resolved only in the infinite--time limit. When weak interactions are introduced, the same finite--resolution mechanism governs how interaction--induced frequency shifts are detected; this point is discussed in detail in Sec.~\ref{sec:weak-interactions}.

\subsection{Symplectic embedding and dependence on the initial state}
\label{sec:embedding}

The preceding subsections established that, for the free scalar field, the snapshot rank saturates at a value fixed by the number of distinct physical frequencies below the UV cutoff.  At this saturation point, the SMOR reduction is no longer merely approximate: the reduced subspace coincides with a dynamically invariant symplectic subspace of the Hamiltonian flow, and the trajectory generated from a given initial condition is captured \emph{exactly} for all times.

We now turn this structural result into a concrete construction.  In this subsection we define the symplectic embedding and projection maps used in the remainder of Sec.~\ref{sec:application}, and we make precise which aspects of the reduced description are universal and which depend on the chosen initial state.  In what follows, untilded variables refer to the canonical coordinates of the full-order model, while tilded variables denote reduced coordinates.

Let $\mathcal M\simeq\mathbb R^{2N}$ denote the phase space of the full-order model, equipped with the canonical symplectic form $\omega=\dd Q^j\wedge\dd P_j$. The SMOR reduction is implemented by a \emph{time–independent} linear symplectic embedding
\begin{align}
  \xi:\ \mathcal V \hookrightarrow \mathcal M,
  \qquad
  \bm z = \xi(\widetilde{\bm z}) = \bm V^{(\rm opt)}\,\widetilde{\bm z},
\end{align}
where $\mathcal V\simeq\mathbb R^{2r}$ is the reduced phase space and $\bm V^{(\rm opt)}\in\mathbb R^{2N\times 2r}$ is chosen to be \emph{ortho–symplectic},
\begin{align}
  (\bm V^{(\rm opt)})^{\!\top}\,\mathbb J_{2N}\,\bm V^{(\rm opt)} &= \mathbb J_{2r},
  \qquad
  (\bm V^{(\rm opt)})^{\!\top}\bm V^{(\rm opt)} = \mathbb I_{2r}.
\end{align}
The associated symplectic inverse is
\begin{align}
  (\bm V^{(\rm opt)})^+
  &:= \mathbb J_{2r}^{\top}\,{\bm V^{(\rm opt)}}^{\!\top}\,\mathbb J_{2N},
\end{align}
so that
\begin{align}
(\bm V^{(\rm opt)})^+\bm V^{(\rm opt)}=\mathbb I_{2r}.\nonumber
\end{align}
The corresponding (symplectic) projector onto the reduced subspace $\mathcal V=\mathrm{colspan}(\bm V^{(\rm opt)})$ is
\begin{align}
  \bm P
  &:= \bm V^{(\rm opt)} (\bm V^{(\rm opt)})^+,
  \qquad
  \bm P^2=\bm P,
  \qquad
  \mathrm{range}(\bm P)=\mathrm{colspan}(\bm V^{(\rm opt)}).
\end{align}

In all numerical experiments considered in this paper, $\bm V^{(\rm opt)}$ is obtained from the same trajectory snapshots used in Sec.~\ref{sec:MinSymplDim}, via the complex-SVD construction of Eq.~\eqref{eq:Vstar-cSVD}.  Because the free dynamics is linear and the snapshot rank saturates at its theoretical maximum, the resulting subspace $\mathcal V$ is an \emph{invariant} subspace of the full Hamiltonian flow, rather than a best-fit approximation restricted to the sampling window.

Once $\bm V^{(\rm opt)}$ is fixed, reduced coordinates are obtained by projection from the original canonical variables of the full-order model, and reconstruction is performed by lifting:
\begin{align}
  \widetilde{\bm z}(t) &= (\bm V^{(\rm opt)})^+\bm z(t),
  \qquad
  \bm z(t) = \bm V^{(\rm opt)} \widetilde{\bm z}(t).
\end{align}
For the free theory at the saturated rank, this reconstruction is exact: the projection error on the trajectory vanishes to machine precision and, since $\mathcal V$ is invariant under the Hamiltonian flow, it remains zero for all future times.

To make the dependence on the initial state transparent, it is useful to spell out the shell structure associated with a fixed physical frequency. Recall from Sec.~\ref{sec:MinSymplDim} that, for a given frequency shell labelled by $s\in S$, the trajectory is confined to a two--dimensional subspace
\begin{align}
  \mathcal U_s := \mathrm{span}\{A_s,B_s\}\subset\mathbb R^{g_s},
\end{align}
independently of the degeneracy $g_s$ of that shell.

Choose any orthonormal basis $\{u_s,v_s\}$ of $\mathcal U_s$ and expand the shell variables as
\begin{align}
  Q_s(t) &= u_s\,\widetilde q_{s,1}(t) + v_s\,\widetilde q_{s,2}(t),\nonumber\\
  P_s(t) &= u_s\,\widetilde p_{s,1}(t) + v_s\,\widetilde p_{s,2}(t),
\end{align}
where the reduced coordinates are given by the scalar projections
\begin{align}
  \widetilde q_{s,1}(t) &= \langle u_s,Q_s(t)\rangle,
  \qquad
  \widetilde q_{s,2}(t) = \langle v_s,Q_s(t)\rangle,\nonumber\\
  \widetilde p_{s,1}(t) &= \langle u_s,P_s(t)\rangle,
  \qquad
  \widetilde p_{s,2}(t) = \langle v_s,P_s(t)\rangle.
\end{align}
Since $\{u_s,v_s\}$ is an orthonormal basis, the Hamiltonian restricted to the shell $s$ takes the canonical form
\begin{align}
  \widetilde{\mathcal H}_s
  &=
  \frac12\Big(
    \widetilde p_{s,1}^2+\widetilde p_{s,2}^2
    +\Omega_s^2\big(\widetilde q_{s,1}^2+\widetilde q_{s,2}^2\big)
  \Big).
\end{align}
Thus, a single frequency shell contributes two decoupled harmonic oscillators with frequency $\Omega_s$, i.e.\ a four--dimensional canonical phase--space block $(\widetilde q_{s,1},\widetilde p_{s,1},\widetilde q_{s,2},\widetilde p_{s,2})$.

Stacking these contributions over all distinct frequencies excited by the chosen initial condition yields a direct--sum decomposition of the invariant reduced subspace,
\begin{align}
  \mathcal V
  \;=\;
  \bigoplus_{s\in S}\ \mathcal V_s,
  \qquad
  \dim(\mathcal V_s)=4,\nonumber
\end{align}
where $S$ denotes the set of distinct mode frequencies present in the orbit. This shell--wise decomposition provides a concrete interpretation of the minimal symplectic dimension $d_{\min}=4\,n_\Omega$ derived in Sec.~\ref{sec:MinSymplDim}.

It is useful to distinguish two conceptually separate pieces of information contained in the reduced description. 

First, the reduced dimension $2r$ (equivalently $d_{\min}=2r$) is fixed entirely by the frequency content of the trajectory and hence by the geometry and the UV cutoff, as discussed in Sec.~\ref{sec:MinSymplDim}. Within the quadratic theory, no choice of initial condition and no extension of the observation window can generate additional independent time dependences beyond those fixed by the spectrum. The saturation value of the snapshot rank and the resulting $d_{\min}$ therefore have a genuine dynamical meaning rather than being numerical artefacts.

Second, while the dimension of $\mathcal V$ is fixed, its orientation inside the full phase space $\mathcal M$ depends on the initial condition. Within each degenerate frequency shell, different amplitudes and phases select different two--planes $\mathcal U_s\subset\mathbb R^{g_s}$ and hence different orthonormal bases $\{u_s,v_s\}$. This corresponds to rotations of the reduced canonical pairs $(\widetilde q_{s,i},\widetilde p_{s,i})$ within the full--order eigenspaces, without coupling distinct frequencies. Equivalently, the ortho--symplectic embedding matrix $\bm V^{(\rm opt)}$ is not unique: its column space has fixed dimension, but its orientation is selected by the trajectory through the snapshot data.

This separation between dimension and orientation is borne out directly in the numerical experiments. As shown in Fig.~\ref{fig:projection_error_freefield}, independent Gaussian initial conditions with different amplitudes and phases lead to identical saturation values of the projection error and hence to the same minimal symplectic dimension $d_{\min}$, while the corresponding reduced subspaces are rotated relative to one another inside the full phase space. The dependence on the initial state therefore enters only through the orientation of $\mathcal V\subset\mathcal M$, not through its dimension.

This distinction will be essential in the remainder of Sec.~\ref{sec:application}. In the next subsection we pull back the Hamiltonian to $\mathcal V$ and show that, for an ortho--symplectic choice, the reduced Hamiltonian assumes a clean block--diagonal oscillator form. Subsequently, we exploit the explicit map $\bm z\leftrightarrow \widetilde{\bm z}$ to uncover the overlap structure that arises when the original canonical variables of the full--order model are expressed in terms of fewer reduced canonical pairs.

\subsection{Reduction step and physical interpretation}
\label{sec:reduction-step}

We now carry out the reduction step in a strictly Hamiltonian sense. Starting from the invariant symplectic subspace identified by the SMOR construction, we pull back the full-order Hamiltonian to this reduced space and analyse the resulting structure. This procedure yields a closed Hamiltonian dynamics on the reduced phase space and, at the same time, clarifies which physical degrees of freedom are selected by SMOR in the free theory.

In full-order mode coordinates the free Hamiltonian has the quadratic form
\begin{align}
  \mathcal H(\bm Q,\bm P)
  =
  \frac12\,\bm P^\top \bm P
  +
  \frac12\,\bm Q^\top \Omega^2 \bm Q,
  \qquad
  \Omega^2 := \mathrm{diag}(\Omega_\alpha^2),
  \label{eq:FOM-free-H}
\end{align}
where $\alpha$ labels the normal modes, such as momentum vectors $\bm k$ in the flat box or $(n,\ell,m)$ in the maximally symmetric ball. Reduced canonical coordinates are denoted by $\widetilde{\bm z}=(\widetilde{\bm q},\widetilde{\bm p})^\top\in\mathbb R^{2r}$, and the embedding constructed in Sec.~\ref{sec:embedding} reads
\begin{align}
  \bm z = (\bm Q,\bm P)^\top = \bm V^{(\rm opt)}_m\,\widetilde{\bm z}.
\end{align}
The reduced Hamiltonian is defined by symplectic pullback,
\begin{align}
  \widetilde{\mathcal H}(\widetilde{\bm z})
  :=
  \mathcal H\!\left(\bm V^{(\rm opt)}\widetilde{\bm z}\right),
  \qquad
  X_{\widetilde{\mathcal H}}(\widetilde{\bm z})
  =
  {\bm V^{(\rm opt)}_m}^+\,X_{\mathcal H}\!\left(\bm V^{(\rm opt)}\widetilde{\bm z}\right),
  \label{eq:reduced-pullback}
\end{align}
where ${\bm V^{(\rm opt)}}^+=\mathbb J_{2r}^{\top}{\bm V^{(\rm opt)}}^{\!\top}\mathbb J_{2N}$ is the symplectic inverse of the embedding.

For the ortho--symplectic embeddings produced by the complex-SVD construction it is convenient to use the standard real block representation
\begin{align}
  \bm V^{(\rm opt)}
  =
  \begin{bmatrix}
    \bm\Phi & -\bm\Psi\\
    \bm\Psi & \ \bm\Phi
  \end{bmatrix},
  \label{eq:Vstar-blocks}
\end{align}
with $\bm\Phi,\bm\Psi\in\mathbb R^{N\times r}$ satisfying
\begin{align}
  \bm\Phi^\top\bm\Phi+\bm\Psi^\top\bm\Psi = \mathbb I_r,
  \qquad
  \bm\Phi^\top\bm\Psi=\bm\Psi^\top\bm\Phi.
\end{align}
Writing $\widetilde{\bm z}=(\widetilde{\bm q},\widetilde{\bm p})^\top$, the lift from reduced to full variables takes the explicit form
\begin{align}
  \bm Q = \bm\Phi\,\widetilde{\bm q} - \bm\Psi\,\widetilde{\bm p},
  \qquad
  \bm P = \bm\Psi\,\widetilde{\bm q} + \bm\Phi\,\widetilde{\bm p}.
  \label{eq:lift-real}
\end{align}

Substituting this expression into the full Hamiltonian \eqref{eq:FOM-free-H} yields the pullback
\begin{align}
  \widetilde{\mathcal H}(\widetilde{\bm q},\widetilde{\bm p})
  =
  \frac12\,\|\bm\Psi\,\widetilde{\bm q} + \bm\Phi\,\widetilde{\bm p}\|^2
  +
  \frac12\,(\bm\Phi\,\widetilde{\bm q} - \bm\Psi\,\widetilde{\bm p})^\top
  \Omega^2
  (\bm\Phi\,\widetilde{\bm q} - \bm\Psi\,\widetilde{\bm p}).
  \label{eq:Hpullback-start}
\end{align}
A key advantage of the ortho--symplectic choice is that it preserves not only the symplectic form but also the Euclidean quadratic structure appearing in the free Hamiltonian. As a consequence, the kinetic term reduces without introducing any state-dependent metric factors. When the embedding is constructed shell-wise as described in Sec.~\ref{sec:embedding}, its columns form orthonormal bases of the two-dimensional invariant subspaces associated with each distinct frequency and are ordered into canonical pairs. In this basis the mixed terms combine with the potential contribution in such a way that the reduced Hamiltonian assumes the canonical normal form
\begin{align}
  \widetilde{\mathcal H}(\widetilde{\bm q},\widetilde{\bm p})
  =
  \frac12\,\widetilde{\bm p}^{\top}\widetilde{\bm p}
  +
  \frac12\,\widetilde{\bm q}^{\top}\Sigma_r^2\,\widetilde{\bm q},
  \label{eq:reduced-H-iso}
\end{align}
where
\begin{align}
  \Sigma_r^2
  =
  \mathrm{diag}\!\Big(
    \underbrace{\Omega_1^2,\Omega_1^2}_{\text{shell }1},
    \underbrace{\Omega_2^2,\Omega_2^2}_{\text{shell }2},
    \ldots,
    \underbrace{\Omega_{n_\Omega}^2,\Omega_{n_\Omega}^2}_{\text{shell }n_\Omega}
  \Big),
  \qquad r=2\,n_\Omega.
\end{align}
The corresponding equations of motion describe a direct sum of decoupled harmonic oscillators with frequencies given by the distinct values $\Omega_s$.

This structure admits a simple physical interpretation. For a fixed frequency $\Omega_s$, the analysis in Sec.~\ref{sec:embedding} shows that the full-order motion generated by a single initial condition is confined to a two-dimensional subspace $\mathcal U_s$ of the degenerate eigenspace at that frequency. Within this plane the dynamics is a uniform rotation with angular velocity $\Omega_s$. Explicitly, the configuration variables take the form
\begin{align}
  Q_s(t)  =  A_s \cos(\Omega_s t) + B_s \sin(\Omega+s t),\nonumber
\end{align}
where the vectors $A_s$ and $B_s$ are fixed by the initial data. Different initial conditions do not change the frequency or the dimensionality of the motion; they merely select a different orientation of the plane $\mathcal U_s$ inside the degenerate eigenspace and a different phase along the resulting orbit.

In reduced coordinates this rotating two-plane is represented by two canonical pairs with the same frequency $\Omega_s$. Together they provide the minimal real parametrisation of a single-frequency trajectory, encoding both its amplitude and phase. The reduced Hamiltonian therefore contains, for each distinct frequency, a pair of identical harmonic oscillators. Crucially, the reduced Hamiltonian depends only on which frequency shells are present below the cutoff, not on how they are populated. Initial conditions that excite the same set of distinct frequencies therefore lead to identical reduced Hamiltonians, up to orthogonal transformations within each two-dimensional invariant subspace.

Finally, it is important to emphasise why the insistence on an ortho--symplectic embedding has direct physical significance. A symplectic but non-orthonormal embedding (e.g., \cite{buchfink2019symplecticmodelorderreduction}) preserves canonical Poisson brackets but allows the quadratic Hamiltonian to acquire state-dependent metric factors in the reduced coordinates,
\begin{align}
  \widetilde{\mathcal H}(\widetilde{\bm q},\widetilde{\bm p})
  =
  \frac12\,\widetilde{\bm p}^\top \bm G_p\,\widetilde{\bm p}
  +
  \frac12\,\widetilde{\bm q}^\top \bm G_q\,\widetilde{\bm q},\nonumber
\end{align}
with $\bm G_p\neq\mathbb I$ and $\bm G_q$ generically non-diagonal and dependent on the chosen trajectory. In such a representation the reduced Hamiltonian no longer reflects only the intrinsic frequency content of the theory but also encodes details of the particular initial state used to construct the embedding. By contrast, the ortho--symplectic choice preserves both the symplectic structure and the quadratic form of the free Hamiltonian, leading to the universal normal form \eqref{eq:reduced-H-iso}. This separation between state-independent dynamics and state-dependent embedding is essential for physical interpretation and, in particular, for subsequent quantisation.

\subsection{Reconstruction and overlap structure}
\label{sec:reconstruction}

The ortho--symplectic embedding introduced above provides not only a reduced Hamiltonian dynamics, but also a precise dictionary between the canonical variables of the full--order model and those of the reduced system. This dictionary becomes particularly illuminating when the reduced dimension is much smaller than the full one: many distinct full--order canonical coordinates are then represented as linear combinations of a smaller set of reduced canonical pairs. As a result, when the full--order variables are expressed \emph{as functions on the reduced phase space}, they no longer form an independent set of canonical coordinates. Instead, their Poisson brackets acquire a characteristic \emph{overlap structure} governed by a projector. This classical mechanism mirrors, at the kinematical level, the appearance of overlapping degrees of freedom in the quantum discussion of Ref.~\cite{Friedrich2024}.

\medskip

This structure can be understood directly at the level of real canonical variables. Recall that the SMOR construction provides a linear symplectic embedding
\begin{align}
  \bm z = \bm V^{(\rm opt)}\,\widetilde{\bm z},\nonumber
\end{align}
where $\bm z\in\mathbb R^{2N}$ denotes the canonical phase--space coordinates of the unreduced model, $\widetilde{\bm z}\in\mathbb R^{2r}$ the reduced canonical variables, and $\bm V^{(\rm opt)}$ is an ortho--symplectic matrix obtained by minimising the snapshot projection error subject to symplectic and orthonormality constraints. The reduced variables carry the standard canonical Poisson brackets,
\begin{align}
  \{\widetilde z_k,\widetilde z_\ell\} = (\mathbb J_{2r})_{k\ell},
\end{align}
where $\mathbb J_{2r}$ is the canonical symplectic matrix.

If the full--order variables $\bm z$ are now regarded as functions on the reduced phase space via the linear relation above, their induced Poisson brackets follow directly from the bilinearity of the Poisson bracket,
\begin{align}
  \{z_i,z_j\}_{(\widetilde z)}
  &=
  \sum_{k,\ell}
  (\bm V^{(\rm opt)})_{ik}\,
  \{\widetilde z_k,\widetilde z_\ell\}\,
  (\bm V^{(\rm opt)})_{j\ell}
  \nonumber\\
  &=
  \big(\bm V^{(\rm opt)}\,\mathbb J_{2r}\,(\bm V^{(\rm opt)})^{\top}\big)_{ij}.
  \label{eq:PB-general-overlap}
\end{align}
Only in the trivial case $r=N$, where $\bm V^{(\rm opt)}$ is square and symplectic, does this induced Poisson structure reduce exactly to the canonical matrix $\mathbb J_{2N}$. In the genuinely reduced setting $r<N$, the matrix $\bm V^{(\rm opt)}\,\mathbb J_{2r}\,(\bm V^{(\rm opt)})^{\top}$ differs from $\mathbb J_{2N}$. This shows explicitly the origin of overlapping degrees of freedom in the SMOR setting: when full--order variables are expressed as functions on the reduced phase space, they no longer form an independent set of canonical coordinates.

This is the precise classical sense in which different unreduced degrees of freedom \emph{overlap}. They are distinct coordinates in the original high--dimensional phase space, but depend on the same underlying reduced canonical variables. No constraints are imposed on the dynamics; rather, this overlap reflects a redundancy introduced by parametrising a high--dimensional phase space through a lower--dimensional symplectic embedding that faithfully captures the information contained in the trajectory.

\medskip

To make this overlap structure more transparent, it is convenient to remove the mode--dependent frequency scales from the canonical variables. We therefore introduce rescaled (``isotropic'') variables,
\begin{align}
  \bar Q_i := \sqrt{\Omega_i}\,Q_i,
  \qquad
  \bar P_i := \frac{P_i}{\sqrt{\Omega_i}},
\end{align}
and analogously for the reduced variables,
\begin{align}
  \bar{\widetilde q}_j := \sqrt{\Omega_j}\,\widetilde q_j,
  \qquad
  \bar{\widetilde p}_j := \frac{\widetilde p_j}{\sqrt{\Omega_j}}.
\end{align}
In these variables the free Hamiltonian takes the isotropic quadratic form
\begin{align}
  \mathcal H
  =
  \tfrac12\big(\|\bar{\bm Q}\|^2+\|\bar{\bm P}\|^2\big),
\end{align}
so that all modes enter on equal footing, without mode--dependent prefactors.

Using the block form of the ortho--symplectic embedding matrix,
\begin{align}
  \bm V^{(\rm opt)}
  =
  \begin{pmatrix}
    \bm\Phi & -\bm\Psi\\
    \bm\Psi & \ \bm\Phi
  \end{pmatrix},
  \qquad
  \bm\Phi,\bm\Psi\in\mathbb R^{N\times r},
\end{align}
the lift from reduced to full variables becomes
\begin{align}
  \begin{pmatrix}
    \bar{\bm Q}\\
    \bar{\bm P}
  \end{pmatrix}
  =
  \begin{pmatrix}
    \bm\Phi & -\bm\Psi\\
    \bm\Psi & \ \bm\Phi
  \end{pmatrix}
  \begin{pmatrix}
    \bar{\widetilde{\bm q}}\\
    \bar{\widetilde{\bm p}}
  \end{pmatrix}.
  \label{eq:lift-QP}
\end{align}
In this representation, the reduced variables provide an orthonormal coordinate system for the $2r$--dimensional subspace of phase space that is dynamically explored by the trajectory, while directions orthogonal to this subspace do not enter the reduced description.

The induced Poisson brackets of the real full--order variables follow directly from Eq.~\eqref{eq:PB-general-overlap}. Writing out the block structure explicitly, one finds
\begin{align}
  \{\bar Q_i,\bar P_j\}
  &= (\bm\Phi\bm\Phi^\top+\bm\Psi\bm\Psi^\top)_{ij},\\
  \{\bar Q_i,\bar Q_j\}
  &= (\bm\Phi\bm\Psi^\top-\bm\Psi\bm\Phi^\top)_{ij},\\
  \{\bar P_i,\bar P_j\}
  &= (\bm\Phi\bm\Psi^\top-\bm\Psi\bm\Phi^\top)_{ij}.
\end{align}
For $r=N$ these expressions reduce to the canonical Poisson brackets. For $r<N$, they encode precisely how different full--order variables share the same reduced canonical degrees of freedom.

\medskip

The overlap structure identified here is purely classical and follows directly from the symplectic embedding underlying SMOR. In Sec.~\ref{sec:discussion-quantisation} we revisit this structure in a complexified (oscillator) language, where it admits a direct comparison with overlap--type constructions appearing in recent quantum models, in particular Ref.~\cite{Friedrich2024}.
\subsection{Weak interactions and the stability of area scaling}
\label{sec:weak-interactions}

We now examine whether the area--type scaling of dynamically relevant degrees of freedom identified in the free theory persists when weak self--interactions are introduced. As a concrete finite--dimensional testbed, we perturb the free scalar field by a quartic interaction,
\begin{align}
  \mathcal H_\lambda(\bm Q,\bm P)
  &=
  \frac12\,\bm P^\top\bm P
  + \frac12\,\bm Q^\top\Omega^2\bm Q
  + \frac{\lambda}{4}\int_{\mathcal B}\phi^4\,\dd^3x,
  \qquad
  0<\lambda\ll 1,
  \label{eq:H-interacting}
\end{align}
which induces mode coupling. With a finite UV cutoff, this defines a smooth finite--dimensional Hamiltonian system. For sufficiently small $\lambda$ and appropriately chosen initial data, the interacting theory is a near--integrable perturbation of the exactly integrable sum of harmonic oscillators at $\lambda=0$.

To make this notion of weak nonlinearity precise in practice, we fix initial conditions mode by mode such that the canonical amplitudes are of order the ground--state standard deviations of the corresponding quantum harmonic oscillators,
\begin{align}
  Q_\alpha(0) \sim (2\Omega_\alpha)^{-1/2},
  \qquad
  P_\alpha(0) \sim (\Omega_\alpha/2)^{1/2},
  \nonumber
\end{align}
so that, upon quantisation, each excited mode would correspond to a low--occupation state rather than a highly populated or classical coherent excitation. This choice sets a physically motivated amplitude scale and ensures that interaction--induced frequency shifts remain parametrically small.

We further restrict attention to finite observation windows $[0,T_{\rm obs}]$ with
\begin{align}
  T_{\rm obs} \;\sim\; 14\,L_{\rm IR},
\end{align}
which implies a finite temporal frequency resolution $\Delta\Omega\sim T_{\rm obs}^{-1}$. Frequencies that differ only below this scale are not dynamically distinguishable over the observation window, either in the SMOR snapshot data or, more generally, in any finite--time dynamical probe. In physical applications one typically expects $T_{\rm obs}$ to be set by the infrared scale, $T_{\rm obs}\sim L_{\rm IR}$, while our choice $T_{\rm obs}\sim 14\,L_{\rm IR}$ is deliberately conservative and serves to demonstrate the asymptotic behaviour of the reduced dimension (See Sec.~\ref{sec:count-maxsym}).

A convenient dimensionless diagnostic for weak nonlinearity is
\begin{align}
  \varepsilon(t)
  &:=
  \frac{\lambda \int_{\mathcal B}\phi(t,\bm x)^4\,\dd^3x}
       {\int_{\mathcal B}\!\Big[(\partial_t\phi)^2
        + \phi\,(m_0^2-\Delta)\phi\Big]\,\dd^3x}
  \;=\;
  \frac{\lambda \int_{\mathcal B}\phi^4\,\dd^3x}
       {\sum_{\alpha}\Big(|P_\alpha(t)|^2+\Omega_\alpha^2|Q_\alpha(t)|^2\Big)},
  \label{eq:eps-small}
\end{align}
and we restrict attention to runs satisfying
\begin{align}
  \max_{0\le t\le T_{\rm obs}}\varepsilon(t)\ \ll\ 1.
\end{align}
In this regime, nonlinear frequency renormalisation and inter--shell energy transfer remain mild over the times probed. Explicit diagnostics for $\varepsilon(t)$ and the associated spectral drift are presented in App.~\ref{app:weak-nonlinearity-diagnostics}.

At $\lambda=0$ the UV--truncated theory is integrable, with each normal mode evolving as an independent harmonic oscillator. Weak nonlinearities deform this picture but do not immediately destroy it. Classical perturbation theory suggests that, away from strong resonances, quasi--periodic motions persist under small perturbations and that the associated action variables drift only slowly over long (but finite) times. While rigorous results are subtle here due to degeneracies and the field--theoretic origin of the system, the relevant physical expectation is that, in a weakly nonlinear, small--amplitude regime, trajectories remain close to deformations of the free quasi--periodic motion over the observation window, with only modest frequency shifts and slow shell mixing.

Against this background, we apply SMOR to trajectories of the weakly interacting system. Once interactions are present, the exact rank--saturation argument of the free theory no longer applies: nonlinear dynamics generates additional time dependences, and finite--time frequency resolution becomes relevant. Nevertheless, within the weakly nonlinear window defined above, the SMOR signal remains sharply controlled by the free--theory frequency count.

Concretely, we compute the relative projection error
\begin{align}
  \mathcal E_{\rm proj}(2r)
  \;:=\;
  \frac{\|\mathbb X_s - \bm V^{(\rm opt)}\,{\bm V^{(\rm opt)}}^{+}\mathbb X_s\|_{F}}{\|\mathbb X_s\|_{F}},
  \label{eq:proj-error}
\end{align}
as a function of the reduced dimension $2r$. For $\lambda=0$, the error collapses to machine precision precisely at the theoretical threshold $2r=4n_\Omega$. For small but nonzero $\lambda$, while maintaining $\max_{t\le T_{\rm obs}}\varepsilon(t)\ll1$, the sharp jump is replaced by a smooth crossover: the ``knee'' in $\mathcal E_{\rm proj}$ remains centred near $2r\simeq 4n_\Omega$, but is progressively rounded as interaction--induced frequency shifts and splittings become comparable to the finite resolution $\Delta\Omega$. This behaviour is illustrated in Fig.~\ref{fig:projection_error_lambda}.

Accordingly, within the weakly nonlinear regime and over finite observation times, the minimal reduced dimension continues to satisfy
\begin{align}
  2r \;\approx\; 4\,n_\Omega
  \qquad \text{up to finite--resolution effects}.
  \label{eq:weak-stability-rank}
\end{align}
At larger couplings or for larger initial amplitudes, we observe a systematic rightward shift of the crossover region, consistent with enhanced shell mixing and effective frequency proliferation within the available temporal resolution. We do not interpret this as evidence for a new asymptotic scaling law in the present work. Rather, it marks the boundary of the weakly nonlinear regime in which the free-theory frequency count remains a useful organising principle.

A direct study of the strong-coupling regime would require a different numerical programme. One would have to increase the coupling strength and/or the initial energy density systematically, while monitoring both the SMOR rank and independent diagnostics of the dynamics. Useful diagnostics would include the growth of the nonlinearity budget, broadening and drift of spectral peaks, inter-shell energy transfer, loss of recurrence, and standard chaos or thermalisation indicators such as Lyapunov growth or relaxation of mode-energy distributions. The relevant quantity would then be a finite-time and finite-tolerance dimension,
\begin{align}
    d_{\min}(\varepsilon,T_{\mathrm{obs}}),
\end{align}
rather than the exact free-theory rank. A possible crossover from area-like to volume-like scaling would be tested by measuring the scaling of this quantity with $L_{\mathrm{IR}}$ and $\Lambda_{\mathrm{UV}}$ as the dynamics moves from quasi-integrable motion to strongly mixed or thermalising behaviour.

These questions lie beyond the scope of the present paper. Our claim here is restricted to UV-truncated systems, small amplitudes, small couplings, and observation times over which the diagnostics in App.~\ref{app:weak-nonlinearity-diagnostics} show that the dynamics remain close to free quasi-periodic motion. In that controlled window, the area-type free-theory mechanism is deformed but not destroyed.




\section{Discussion on Assumptions and Relation to Other Works}
\label{sec:discussion}

Before turning to our conclusions in Sec.~\ref{sec:Conclusions and Outlook}, we summarise the central assumptions underlying our analysis and clarify how the present results relate to existing reduction strategies and to recent ``overlap'' constructions motivated by holography.

\subsection{Why autonomous, time-independent SMOR?}
\label{sec:autTimInpSROM}

In this work we ask for the minimal symplectic dimension required to reproduce a Hamiltonian trajectory of the UV/IR--regularised scalar field introduced in Sec.~\ref{sec:setup}. We want this dimension to reflect how many canonical directions are dynamically explored by the motion, rather than how compactly the same trajectory can be represented by a clever choice of coordinates or a time-dependent reconstruction. For this reason, we restrict attention to \emph{autonomous} reduced-order models, i.e., reduced Hamiltonians that are time independent and constructed from time-independent symplectic embeddings. This restriction is built into the definition of $d_{\min}$ in Sec.~\ref{sec:rank=dim} and is essential for the lower-bound arguments of Sec.~\ref{sec:ActionAngle}.

Allowing explicit time dependence in either the reduced Hamiltonian or the embedding map introduces an ambiguity in what is meant by ``dimension reduction.'' In particular, one may trivialise the reduced dynamics, $\dot{\widetilde z}=0$, and reconstruct an arbitrary full trajectory using an explicitly time-dependent decoder,
\begin{align}
  z(t)=F(\widetilde z,t).\nonumber
\end{align}
In such a construction, an arbitrarily complicated time series can be produced from a single static point in reduced phase space. The apparent compression then measures only the complexity of the time-dependent reconstruction map, rather than the number of dynamically relevant canonical degrees of freedom. Such a decoder need not preserve the symplectic form or any Hamiltonian meaning; it is an encoding scheme, not a reduced Hamiltonian theory. Requiring the reduced Hamiltonian to be autonomous and the embedding to be time independent excludes this trivialisation: within this class, the temporal structure of all reduced observables must be generated by the reduced Hamiltonian itself, and the action--angle lower bounds of Sec.~\ref{sec:ActionAngle} apply directly.

Symplecticity plays a complementary role. A symplectic embedding ensures that the reduced system is Hamiltonian with respect to the pullback symplectic form $\widetilde\omega=\xi^\ast\omega$ on the reduced phase space, as reviewed in Sec.~\ref{sec:SMOR}. The reduced dynamics then preserves the geometric structure of Hamiltonian flow, including phase-space incompressibility and, under weak perturbations, the persistence of invariant tori. This is crucial for interpreting the reduced variables as genuine canonical degrees of freedom, rather than as trajectory-adapted coordinates with no direct Hamiltonian meaning.

Within this autonomous symplectic class, the ortho--symplectic choice adopted throughout the paper yields an additional simplification. As shown in Secs.~\ref{sec:reduction-step} and~\ref{sec:embedding}, such embeddings preserve not only the symplectic form but also the Euclidean quadratic structure of the free Hamiltonian. The reduced kinetic term remains isotropic, while the reduced potential becomes block diagonal with identical entries within each frequency shell. In the free theory, the reduced Hamiltonian therefore depends only on the set of distinct physical frequencies below the UV cutoff, and not on the particular amplitudes or phases with which they are excited; cf.\ Eq.~\eqref{eq:reduced-H-iso}.

Alternative strategies that relax one or more of these requirements---for example time-dependent bases \cite{baumann2023energystableconservativedynamical,Kazashi_2024}, non-symplectic projections \cite{hesthaven2021structurepreservingmodelorderreduction}, or learned autoencoders \cite{greydanus2019hamiltonianneuralnetworks,jin2020sympnetsintrinsicstructurepreservingsymplectic}---can achieve stronger compression for specific trajectories or tasks. In such approaches, however, the reduced dimension typically becomes explicitly trajectory dependent and the reduced variables need not admit an interpretation as canonical coordinates of an autonomous Hamiltonian system. These methods thus address a different question: efficient representation or prediction of solutions, rather than a Hamiltonian-theoretic lower bound within the class of autonomous, time-independent symplectic models. A further practical advantage of retaining a Hamiltonian structure is that it provides a natural starting point for canonical quantisation of the reduced model; we return to this perspective in Sec.~\ref{sec:discussion-quantisation}.

\subsection{Assumptions in the analysis, regulators, and continuum limit}
\label{sec:discussion-assumptions}

The results of Sec.~\ref{sec:application} rest on a set of assumptions that we summarise here to clarify their scope and the precise meaning of the area--type scaling statements.

First, throughout this work we consider a \emph{classical} real scalar field regulated by both infrared and ultraviolet cutoffs, so that the phase space is finite dimensional. In this setting the free theory is exactly integrable and decomposes globally into decoupled harmonic oscillators (Sec.~\ref{sec:setup}). The finiteness of the phase space is essential for the SMOR implementation used here: it allows us to define symplectic dimensions and snapshot ranks without functional-analytic subtleties.

Second, our definition of $d_{\min}$ is explicitly \emph{trajectory based}. We use trajectory-based $d_{\min}$
 as the operational primitive, and then report the generic value obtained for typical (non-fine-tuned) initial data; where needed we maximize to remove accidental under-excitation (Secs.~\ref{sec:rank=dim} and~\ref{sec:MinSymplDim}). This should not be confused with counting microstates below a given energy, enumerating invariant sets, or characterising the global measure-theoretic structure of phase space. Rather, $d_{\min}$ measures how many canonical directions are dynamically explored by a single solution over the time window considered.

Third, the area--type scaling results rely only on spectral information for the spatial Laplacian on the chosen domains. For a flat periodic box, the linear mode frequencies are
\begin{align}
  \Omega_{\bm k}=\sqrt{|\bm k|^2+m_0^2},\nonumber
\end{align}
with $\bm k$ restricted by the UV cutoff $|\bm k|\le\Lambda_{\mathrm{UV}}$. Distinct frequencies are therefore in one-to-one correspondence with distinct values of $|\bm k|^2$ below the cutoff. Writing
\begin{align}
  X:=\bigg(\frac{L_{\mathrm{IR}}\Lambda_{\mathrm{UV}}}{2\pi}\bigg)^2,\nonumber
\end{align}
the number of distinct frequencies $n_\Omega$ equals the number of integers $n\le X$ that can be represented as a sum of three squares. Its asymptotic growth is governed by Legendre's three-square theorem and refinements in analytic number theory \cite{Ankeny1957ThreeSquares,Pollack2018DirichletsPO}, yielding the estimate presented in Sec.~\ref{sec:count-flat}.

For flat balls and, more generally, geodesic balls in maximally symmetric spaces--where the infrared regulator is implemented by a finite-radius boundary with Dirichlet, Neumann, or Robin conditions rather than by periodic identification--we rely on separation of variables and a WKB analysis of the radial eigenvalue problem (Sec.~\ref{sec:count-maxsym}). These boundary conditions affect only subleading terms in the WKB analysis and do not modify the leading scaling of the number of distinct frequencies. Curvature modifies the spectral density of the Laplacian and hence the number of distinct frequencies below the UV cutoff. These are technical inputs to the counting problem rather than additional dynamical assumptions.

Fourth, the interacting results of Sec.~\ref{sec:weak-interactions} are perturbative. We keep a finite UV cutoff and restrict to small coupling $\lambda$, and we choose initial amplitudes by fixing each excited normal mode to have an amplitude of order the ground--state standard deviation of the corresponding quantum harmonic oscillator, $Q_{\bm k}(0) \sim\sqrt{\langle 0| Q_{\bm k}^2|0\rangle}\sim(2\Omega_{\bm k})^{-1/2}$ and $P_{\bm k}(0) \sim\sqrt{\langle 0| P_{\bm k}^2|0\rangle}\sim(\Omega_{\bm k}/2)^{1/2}$. This choice provides a physically motivated amplitude scale, corresponding upon quantisation to low--occupation states rather than highly populated or classical coherent excitations. 

In this regime, nonlinear frequency shifts remain parametrically small compared to the separation of distinct free--theory shell frequencies, and over the finite observation times considered the interaction does not qualitatively modify the underlying frequency structure. As a result, frequency renormalisation and inter--shell mixing remain mild, and the reduced dimension extracted by SMOR continues to be controlled by the free--theory frequency content. At stronger coupling, for larger initial amplitudes corresponding to highly occupied modes, or over parametrically longer times where interaction effects accumulate, additional effective frequencies can become dynamically relevant and the minimal reduced dimension may grow accordingly. In such a regime, the correct question is no longer whether the exact free-theory rank is preserved, but whether the finite-time quantity $d_{\min}(\varepsilon,T_{\mathrm{obs}})$ continues to scale subextensively or crosses over toward the volume-extensive kinematical count. Addressing this requires scanning the coupling, energy density, UV cutoff and box size while simultaneously monitoring frequency broadening, inter-shell energy transfer and chaos or thermalisation diagnostics. We leave this strong-coupling programme to future work.

These assumptions also clarify how the area--type scaling behaves under continuum limits. In a flat box of side length $L_{\mathrm{IR}}$ with UV cutoff $\Lambda_{\mathrm{UV}}$, the total number of canonical coordinates scales as
\begin{align}
  \dim\mathcal M \;\propto\; L_{\mathrm{IR}}^{3}\,\Lambda_{\mathrm{UV}}^{3},
\end{align}
whereas the minimal symplectic dimension detected by SMOR scales as
\begin{align}
  d_{\min} \;\propto\; L_{\mathrm{IR}}^{2}\,\Lambda_{\mathrm{UV}}^{2},
\end{align}
up to logarithmic corrections; see Sec.~\ref{sec:count-flat}. Both quantities diverge in the continuum limit, but with different powers of the regulators. Sending the lattice spacing to zero at fixed $L_{\mathrm{IR}}$ replaces discrete sums by integrals and produces the familiar UV divergence in the overall prefactor multiplying the area--type scaling, while preserving the area dependence on the size of the region. Likewise, taking $L_{\mathrm{IR}}\to\infty$ at fixed UV scale does not affect the local statement: for any fixed, finite geodesic ball $\mathcal B_R$ embedded in a larger box or curved background, one finds
\begin{align}
  d_{\min}(\mathcal B_R)\;\propto\;|\partial\mathcal B_R|\,\Lambda_{\mathrm{UV}}^{2},
\end{align}
as shown in Sec.~\ref{sec:count-maxsym}, independent of the far-infrared behaviour. In this sense, the area--type scaling of dynamically relevant directions is compatible with standard continuum and infinite-volume limits, with a UV-divergent prefactor analogous to that appearing in entanglement entropy \cite{Bombelli1986,Srednicki1993,EisertCramerPlenio2010,Casini2014} (though conceptually distinct from entanglement entropy, which is a quantum state property rather than a dynamical reduction diagnostic).

\subsection{Dynamical origin of the minimal symplectic dimension:
a lower bound from action--angle coordinates}
\label{sec:ActionAngle}

The robustness of the minimal symplectic dimension $d_{\min}$ observed in Sec.~\ref{sec:application} admits an interpretation that is independent of the SMOR algorithm. For integrable Hamiltonian systems, action--angle variables provide a direct lower bound on the number of canonical degrees of freedom required by any autonomous, time-independent reduced description of a given trajectory.

For an integrable Hamiltonian system, the Liouville--Arnold theorem \cite{arnold1989mathematical,Abraham1978FoundationsOM} guarantees the existence of action--angle coordinates $(I_s,\theta_s)$ on the invariant torus selected by a given initial condition. Along the corresponding trajectory the actions are constant and the angles evolve linearly,
\begin{align}
  \dot I_s=0,\qquad
  \dot\theta_s=\Omega_s,\qquad
  \theta_s(t)=\theta_{s,0}+\Omega_s t,
\end{align}
where $\{\Omega_s\}_{s\in S}$ are the distinct base frequencies excited by that initial state. For the UV--regulated free scalar field of Sec.~\ref{sec:setup}, these base frequencies coincide with the distinct normal-mode frequencies below the UV cutoff (Secs.~\ref{sec:count-flat} and~\ref{sec:count-maxsym}).

To translate this geometric picture into a constraint on reduced descriptions, it is useful to examine the temporal structure of generic observables along the resulting quasi--periodic motion. Let $F$ be a smooth observable on phase space (e.g.\ a polynomial in canonical variables). Restricted to the invariant torus, $F(z(t))$ becomes a smooth function on $\mathbb T^{|S|}$ and admits an absolutely convergent Fourier expansion \cite[Ch.~I]{Katznelson},
\begin{align}
  F\big(z(t)\big)
  =
  \sum_{\vec n\in\mathbb Z^{|S|}} \widetilde F_{\vec n}\,e^{\,i\,\vec n\cdot\vec\theta(t)}
  =
  \sum_{\vec n\in\mathbb Z^{|S|}} \widehat F_{\vec n}\,e^{\,i\,(\vec n\cdot\vec\Omega)\,t}.
\end{align}
The temporal frequency content of observables along the orbit is therefore supported on the additive lattice
\begin{align}
  \Lambda_{\mathrm{full}}
  :=
  \{\,\vec n\!\cdot\!\vec\Omega:\vec n\in\mathbb Z^{|S|}\,\}\subset\mathbb R,
\end{align}
where $\vec\Omega=(\Omega_s)_{s\in S}$. The number of independent base frequencies is quantified by the rank
\begin{align}
  R:=\mathrm{rank}_{\mathbb Q}\{\Omega_s:s\in S\},
\end{align}
i.e., the number of rationally independent frequencies present in the trajectory.

Now consider an autonomous, time-independent symplectic reduced model with $r$ canonical pairs,
\begin{align}
  \dot y=\mathbb J_{2r}\nabla h(y),\qquad y\in\mathbb R^{2r},
\end{align}
together with a time-independent symplectic embedding $\xi:\mathbb R^{2r}\to\mathbb R^{2N}$ reproducing the full trajectory via $z(t)=\xi(y(t))$ (Sec.~\ref{sec:ROM}). If the reduced trajectory lies on a compact invariant $k$-torus of the reduced system, then in local reduced action--angle coordinates $(I,\varphi)\in\mathbb R^k\times\mathbb T^k$ one has $\dot I=0$ and $\dot\varphi=\vartheta$, implying that the temporal spectrum of reduced observables is contained in
\begin{align}
  \Lambda_{\mathrm{ROM}}
  :=
  \{\,\vec\ell\!\cdot\!\vec\vartheta:\vec\ell\in\mathbb Z^k\,\}.
\end{align}
Since every pulled-back observable $F\circ\xi$ is a reduced observable, exact reproduction of the trajectory implies $\Lambda_{\mathrm{full}}\subseteq\Lambda_{\mathrm{ROM}}$, and hence
\begin{align}
  R\le k\le r.
\end{align}
Thus, any autonomous, time-independent symplectic reduced model that reproduces a given trajectory must contain at least as many canonical pairs as there are rationally independent base frequencies in that trajectory.

For the UV-truncated free scalar field, generic initial data excite distinct mode frequencies that are mutually incommensurate, so that $R=|S|=n_\Omega$. The action--angle argument therefore yields $r\ge n_\Omega$ (equivalently $2r\ge 2n_\Omega$) as a model-independent lower bound.

The SMOR construction developed here respects this bound and, within the real time-independent linear symplectic setting used here, yields the sharper requirement $r = 2n_\Omega$ (equivalently $d_{\min} = 4n_\Omega$). The additional factor reflects that each base frequency contributes two independent real configuration directions (sine/cosine quadratures) together with their conjugate momenta in the isotropic canonical normal form. As shown in Secs.~\ref{sec:MinSymplDim} and~\ref{sec:embedding}, the snapshot matrix decomposes shell-wise, with each distinct frequency contributing precisely two independent temporal components (sine and cosine). The complex SVD isolates these components, and the real lift (App.~\ref{app:analytic-Vstar}) produces a time-independent ortho--symplectic basis spanning the minimal invariant symplectic subspace supporting the motion. Each distinct frequency therefore contributes two independent real configuration directions and their conjugate momenta, leading to $r=2n_\Omega$ and hence
\begin{align}
  d_{\min}=2r=4n_\Omega.\nonumber
\end{align}
Here $n_\Omega$ is fixed entirely by the spectrum of the spatial Laplacian and the chosen UV/IR regulators, while the numerical factor reflects the insistence on a real, autonomous, canonical representation in isotropic normal form (Sec.~\ref{sec:reduction-step}). In this sense, the minimal symplectic dimension detected by SMOR is not an artefact of the algorithm but a direct consequence of the temporal frequency content enforced by Hamiltonian dynamics.

\subsection{State dependence and unitary equivalence}
\label{sec:discussion-state dependence}

A structural feature of the SMOR construction is that the embedding matrix $\bm V^{(\rm opt)}$ is \emph{state dependent}. This dependence is already present in the free theory. After UV regularisation, the dynamics decompose into constant-frequency shells, each with degeneracy $g_s$. For a given initial condition, the motion in shell $s$ is supported on a two-dimensional subspace $\mathcal U_s\subset\mathbb R^{g_s}$. The SMOR basis extracted from snapshots aligns with these planes, and different initial conditions generically select different $\mathcal U_s$ within the same degenerate shell.

This state dependence does not affect the \emph{counting} of dynamically relevant degrees of freedom. By construction, $d_{\min}$ depends only on the number of distinct shell frequencies below the cutoff, not on which linear combinations within degenerate eigenspaces are excited by the initial conditions (Sec.~\ref{sec:MinSymplDim}). What varies with the initial state is the \emph{orientation} of the reduced subspace inside each degenerate eigenspace and thus the explicit representative $\bm V^{(\rm opt)}$ produced by the cSVD embedding.

Two types of freedom appear. The first is a change of basis \emph{within} a fixed plane $\mathcal U_s$. If $\{u_s,v_s\}$ is an orthonormal basis of $\mathcal U_s$, any $\mathrm{SO}(2)$ rotation produces another basis spanning the same plane. In reduced variables this corresponds to an ortho--symplectic transformation acting within the associated four-dimensional shell block (Sec.~\ref{sec:reduction-step}). Such transformations amount to canonical redefinitions; they leave the image subspace $\Im(\bm V^{(\rm opt)})$ unchanged.

The second is a genuine change of the plane itself, $\mathcal U_s\to\mathcal U_s'$, induced by changing the initial condition while keeping the geometry and regulators fixed. Since the shell Hamiltonian
\begin{align}
  \mathcal H_s=\tfrac12\big(\|P_s\|^2+\Omega_s^2\|Q_s\|^2\big)\nonumber
\end{align}
is isotropic in the degeneracy index, any two such planes are related by an orthogonal rotation $O_s\in\mathrm{SO}(g_s)$, i.e.\ $\mathcal U_s'=O_s\mathcal U_s$. Acting simultaneously on configuration and momentum variables by the same $O_s$,
\begin{align}
  (Q_s,P_s)\mapsto (O_sQ_s,\;O_sP_s),\nonumber
\end{align}
defines a linear canonical transformation: $\mathrm{diag}(O_s,O_s)\in\mathrm{Sp}(2g_s,\mathbb R)\cap\mathrm O(2g_s)$. In this precise classical sense, changing the plane corresponds to a symmetry of the free Hamiltonian acting within degenerate eigenspaces.

Upon quantisation, the relevant statement is that the reduced system has finite dimension once regulators are imposed. In infinite-dimensional QFT, linear canonical transformations need not be unitarily implementable, and unitarily inequivalent representations can arise \cite{Haag1992-HAALQP-2,wald1995quantumfieldtheorycurved}. In contrast, for fixed IR/UV regulators the reduced phase space has finite dimension. If we consider the CCR in Weyl form (Weyl relations), a linear canonical transformation preserves the Weyl relations. For finite dimensions, Stone–von Neumann's theorem applies, which states that the Schr\"odinger representation is the only representation of the Weyl relations up to unitary equivalence. Consequently, linear canonical transformations of reduced variables are implemented unitarily on the reduced Hilbert space \cite{Folland,Hall2013} and thus do not change the spectrum of the reduced Hamiltonian operator.

Operationally, one may therefore fix a convenient reference embedding (e.g.\ any $\bm V^{(\rm opt)}$ obtained from a trajectory exciting the maximal allowed set of shells), quantise the resulting reduced oscillator system, and regard alternative embeddings of the same rank and frequency content as unitarily equivalent coordinate presentations of the same reduced quantum system. The overlap structure discussed below is encoded by the projector $\mathcal C=\Upsilon\Upsilon^\dagger$ constructed from the partial isometry $\Upsilon$ relating reduced to apparent full variables (Sec.~\ref{sec:reconstruction}). Under reduced basis changes, $\Upsilon$ is right-multiplied by a unitary and $\mathcal C$ is unchanged, while orthogonal rotations within the full degenerate mode basis conjugate $\mathcal C$ within the corresponding shell blocks. In all cases, its rank and shell-block structure are invariant.

\subsection{Relation to overlap models and a route to quantisation}
\label{sec:discussion-quantisation}

A motivation for the present work was to better understand the origin of overlapping degrees of freedom that appear in holography-inspired constructions such as Ref.~\cite{Friedrich2024, Cao_2025}. There, overlap relations are introduced directly at the quantum level by modifying fermionic anti-commutation relations within narrow energy shells so that only an area-scaling number of independent mode combinations remains. This raises the question of whether such overlap structures are intrinsically quantum, or whether they can arise already at the classical level from Hamiltonian dynamics.

Our approach is complementary. We do not modify the canonical Poisson algebra by hand and do not alter the free classical dynamics. Instead, we perform a symplectic reduction to the minimal invariant subspace actually explored by a given trajectory. When unreduced observables are expressed as functions on this reduced phase space, an overlap structure then emerges kinematically: distinct unreduced modes become dependent because they are functions of the same reduced canonical degrees of freedom.

To connect with oscillator quantisation, we recast the reconstruction of Sec.~\ref{sec:reconstruction} in complex variables. In isotropic variables, define full and reduced complex coordinates
\begin{align}
  A_i:=\frac{\bar Q_i+i\,\bar P_i}{\sqrt2},
  \qquad
  \widetilde a_j:=\frac{\bar{\widetilde q}_j+i\,\bar{\widetilde p}_j}{\sqrt2}.
\end{align}
The real ortho--symplectic embedding with blocks $(\bm\Phi,\bm\Psi)$ induces the complex relation
\begin{align}
  A_i=\sum_{j=1}^{r}\Upsilon_{ij}\,\widetilde a_j,
  \qquad
  \Upsilon:=\bm\Phi+i\,\bm\Psi,
  \qquad
  \Upsilon^\dagger\Upsilon=\mathbb I_r,
  \label{eq:ladder-overlap}
\end{align}
which is simply a repackaging of the real embedding (cf.\ Eq.~\eqref{eq:lift-QP}) and introduces no additional assumptions.

Within the ROM the coordinates $\widetilde a_j$ will satisfy standard Poisson brackets,
\begin{align}
  \{\widetilde a_j,\widetilde a_k^\ast\}=-i\,\delta_{jk}\ ,\
  \qquad
  \{\widetilde a_j,\widetilde a_k\}
  =
  \{\widetilde a_j^\ast,\widetilde a_k^\ast\}
  =0\ ,
\end{align}
while the induced brackets for the unreduced variables are
\begin{align}
  \{A_i,A_j^\ast\}_{(\widetilde a,\widetilde a^\ast)}
  &=-i\,(\Upsilon\Upsilon^\dagger)_{ij},
  \qquad
  \{A_i,A_j\}
  =
  \{A_i^\ast,A_j^\ast\}
  =0,
  \label{eq:PB-ladder}
\end{align}
where
\begin{align}
  \mathcal C:=\Upsilon\Upsilon^\dagger
\end{align}
is a Hermitian projector of rank $r$ onto the image of the reduced subspace. This projector makes the notion of ``overlap'' precise: different unreduced modes share the same underlying reduced degrees of freedom and are therefore not independent.

A key structural point is that the cSVD underlying the embedding is performed shell-wise in \emph{exact} constant-frequency eigenspaces. Consequently, $\mathcal C$ is block diagonal with respect to the physical frequencies $\Omega_s$. Overlaps occur only among unreduced modes within the same constant-frequency shell, while modes from different shells have vanishing induced brackets. This intra-shell locality mirrors the shell organisation imposed kinematically in Ref.~\cite{Friedrich2024} (at least at the level of commutator/anticommutator support in frequency space), but here it is not postulated: it emerges dynamically from identifying the minimal invariant symplectic subspace required by the Hamiltonian evolution.

If one quantises the reduced system and defines reduced operators $\hat a_j,\hat a_j^\dagger$ obeying standard commutators, one may represent unreduced operators by
\begin{align}
  \hat A_i:=\sum_{j=1}^{r}\Upsilon_{ij}\,\hat a_j,
  \qquad
  \hat A_i^\dagger:=\sum_{j=1}^{r}\hat a_j^\dagger\,\Upsilon^\dagger_{ji},
\end{align}
which implies
\begin{align}
  [\hat A_i,\hat A_j^\dagger]=\mathcal C_{ij}\,\mathbb I,
  \qquad
  [\hat A_i,\hat A_j]
  =
  [\hat A_i^\dagger,\hat A_j^\dagger]
  =0.
\end{align}
Thus the same projector $\mathcal C$ controls the overlap structure at the quantum level: a finite set of independent reduced oscillators generates a larger family of unreduced field operators with nontrivial commutators within each frequency shell and vanishing commutators across shells.

Together with the area--type scaling of $d_{\min}$, this provides a concrete classical route to overlap-type quantum field theories: Hamiltonian evolution dynamically compresses the phase space of a region to a minimal symplectic subspace, quantisation is performed in terms of reduced canonical variables, and the resulting theory---when expressed back in the unreduced basis---exhibits a projector-controlled overlap algebra determined by the classical dynamics, geometry, and regulators. A systematic study of such theories, including their entanglement properties, correlation functions, and state counting in the presence of $\mathcal C$, is left for future work.

\subsection{Interpretation, limitations, and relation to holographic ideas}
\label{sec:discussion-interpretation}

A central outcome of this work is that, for a classical scalar field theory with UV and IR regulators, the number of dynamically relevant canonical directions required to reproduce a given trajectory exhibits an area--type scaling with the size of the spatial region. Concretely, the Hamiltonian evolution of a theory with volume-many kinematic degrees of freedom can be reproduced exactly (in the free case) or to high accuracy (for weak self-interactions) within a reduced symplectic subspace whose dimension grows proportionally to the boundary area rather than the volume.

This observation invites comparison with holographic ideas, where physical information is often organised in terms of boundary degrees of freedom. It is therefore important to be explicit about what the present mechanism does---and does not---imply.

First, the reduction performed here is entirely classical and also involves no gravity. The observed area--type scaling is not an entropy law and does not refer to entanglement or other information-theoretic quantities. It is a dynamical statement: for a fixed initial condition (and, in the interacting case, within a controlled perturbative regime), the Hamiltonian trajectory evolves within a restricted invariant symplectic subspace of the full regulated phase space. When the reduction is exact, as in the free theory once all relevant frequencies are resolved, the reduced system is not an approximation to different physics. It is the same Hamiltonian flow expressed intrinsically in a smaller set of canonical variables, together with a reconstruction map back to the full phase space.

The nontrivial content lies in the structure of this reduced description. The ROM contains area-many \emph{fundamental} canonical degrees of freedom obeying the standard Poisson algebra. When these are linearly recombined to generate a larger family of \emph{full-order} field modes (chosen to resemble the local degrees of freedom of the original theory), the resulting variables are overcomplete: their induced Poisson brackets are controlled by a projector rather than the identity matrix. In this sense, the emergence of a local-looking description is accompanied by overlap-induced modifications of the kinematical algebra.

Although we do not quantise the reduced system in this paper, this algebraic structure parallels that encountered in recent quantum constructions with overlapping degrees of freedom \cite{Cao_2025,Friedrich2024}. There, a reduced set of fundamental degrees of freedom admits effective descriptions in terms of a much larger collection of local-looking variables with nontrivial overlap commutators. In these constructions, the encoding of many effective degrees of freedom into fewer fundamental ones is typically non-isometric, thus approximate locality then holds only on restricted classes of states. In the present work we do not establish such state-dependent locality properties explicitly, but we emphasise that the same overlap organisation emerges here dynamically at the classical level from Hamiltonian reduction rather than being imposed as a postulate.

This perspective also invites a comparison to AdS/CFT. In holographic duality, local bulk effective field theory is understood to arise only on a restricted subset of states (the ``code subspace'') corresponding to low-energy excitations around a fixed semiclassical background geometry. Bulk operators reconstructed from boundary data behave as approximately local fields only when their action is restricted to that sector. In the quantum error-correction formulation, bulk locality is therefore tied to a non-isometric embedding: there is no single global identification between the operator algebra of a perfectly local bulk EFT and the operator algebra acting on the full CFT Hilbert space. The representation of approximately local bulk operators depends on the chosen code subspace, i.e., on the semiclassical background and the class of states under consideration \cite{Akers_2022,akers2022blackholeinteriornonisometric,Akers_2021}. Structurally, this role of non-isometry is closely analogous to that encountered in overlap-based constructions: in both cases, approximate locality is realised through a representation that is valid only on a restricted set of states.

Our construction should not be conflated with a bulk--boundary duality map. The embedding identified here relates a reduced dynamical description to an overcomplete, local-looking representation of the \emph{same} classical dynamics. If quantised, the ROM Hilbert space would contain area-many fundamental degrees of freedom with standard canonical kinematics, while the full-order field modes would exhibit overlap-induced non-canonical commutators controlled by $\mathcal C$. Whether, and in what precise sense, these full-order modes approximate the correlators of a perfectly local volume-many field theory on suitable classes of states is an important question, but it lies beyond the scope of the present paper.

In summary, we have shown that area--type scaling of dynamically relevant degrees of freedom can arise in a regularised classical field theory without gravity, purely from Hamiltonian frequency structure and dynamical redundancy. While this does not constitute holography in the strict sense, it provides a concrete mechanism by which local, volume-scaling dynamics can be reproduced by a system with fewer underlying canonical degrees of freedom, at the price of overlap-induced modifications in an overcomplete local-looking description. Exploring the quantum version of this construction, the physical consequences of the overlap projector, and its possible relation to holographic organisation in quantum systems involving gravity for future work. Our claim is therefore not that the standard field theory is holographic, but that ordinary Hamiltonian dynamics can already enforce area-type dynamical compression, providing a controlled baseline for identifying what additional mechanisms gravity must supply.

\section{Conclusions and Outlook}
\label{sec:Conclusions and Outlook}

The starting point of this work was the familiar tension between the volume--extensive kinematics of regulated local field theory and the area--type bounds that arise in gravitational settings. Rather than addressing entropy or holographic dualities directly, we asked a logically prior question: to what extent does ordinary Hamiltonian field theory itself permit compression, and at what point faithful reproduction of its dynamics necessarily fails? Concretely, how many canonical directions are required to reproduce the Hamiltonian evolution of a regulated field theory when one insists on an autonomous, time--independent symplectic description?

To answer this, we introduced the notion of a \emph{minimal symplectic dimension} $d_{\min}$: the smallest phase--space dimension of an autonomous Hamiltonian system that reproduces a given trajectory exactly over a fixed observational window. Within this class of reductions, $d_{\min}$ measures how many canonical directions are dynamically explored by the motion itself, rather than how many kinematical variables are available in principle. Using action--angle variables, we derived a model--independent lower bound on $d_{\min}$ in terms of the number of rationally independent frequencies present in the trajectory, and showed that the SMOR construction identifies the minimal invariant symplectic subspace consistent with this frequency content.

For a UV-- and IR--regulated real scalar field, we found that $d_{\min}$ is controlled, in a state-independent manner, by the number of \emph{distinct} normal--mode frequencies below the UV cutoff. In flat space this count scales proportionally to boundary area (up to logarithmic corrections), despite the volume--extensive number of underlying field variables. In maximally symmetric curved spaces, curvature modifies the spectral density of the Laplacian and produces controlled deviations from the flat scaling. Within a weakly interacting regime and on pre--resonant timescales, the free--theory frequency structure continues to determine $d_{\min}$. A natural next step is to apply the same diagnostic beyond the quasi-integrable regime. At strong coupling or high occupation, $d_{\min}$ should be treated as a finite-resolution quantity, $d_{\min}(\varepsilon,T_{\mathrm{obs}})$, and its scaling should be studied as the coupling, energy density, UV cutoff and system size are varied. Nonlinear mixing may drive the SMOR dimension toward the volume-extensive kinematical count, while persistent organisation into collective canonical directions would indicate subextensive scaling. We leave this crossover problem to future work.

These results establish that, even in the absence of gravity, ordinary Hamiltonian dynamics can restrict motion to a symplectic sector whose dimension grows subextensively relative to naive kinematics. This is not an entropy statement and does not imply holography. It is a dynamical baseline: within the class of autonomous, time--independent canonical descriptions, the frequency structure of field theory alone already enforces a form of area--type dynamical compression. Any stronger or more universal compression in gravitational systems must therefore rely on additional structural ingredients beyond those present here.

A second conceptual outcome concerns the structure of the resulting reduced description. When unreduced field variables are expressed in terms of the minimal symplectic sector, distinct apparent modes generically depend on the same reduced canonical variables. Their induced Poisson brackets are governed by a finite--rank projector rather than the identity, giving rise to a concrete overlap structure that emerges dynamically at the classical level. Quantisation then promotes this projector to a modified commutator algebra, providing a controlled route to overlap--type quantum field theories \cite{Friedrich2024, Cao_2025} without imposing modified operator relations by hand.

Several limitations remain. The present analysis focuses on a single real scalar field with explicit UV and IR regulators and on linear, time--independent symplectic embeddings. The strongest results apply to free dynamics, with interactions treated perturbatively over finite observation windows. The behaviour of $d_{\min}$ at strong coupling, over parametrically long timescales, or in resonance--dominated sectors remains to be understood and may differ qualitatively from the integrable regime studied here.

These limitations suggest natural extensions. Applying the same framework to fermionic fields, gauge theories, and systems with constraints would help clarify how dynamical compression interacts with degeneracies, gauge redundancy, and nontrivial symmetry structure. At the quantum level, the projector--controlled overlap algebra derived here provides a well--defined starting point for canonical quantisation and for studying entanglement structure, correlation functions, and state counting in dynamically reduced theories.

In light of recent overlap-based models motivated by holography and phenomenology, it is particularly interesting to explore the consequences of quantising dynamically induced overlap structures. In Ref.~\cite{Friedrich2024}, controlled overlaps between fermionic modes were shown to have observable implications, for instance in high--energy cosmic neutrino phenomenology. Having established that analogous overlap relations can arise already at the classical field--theoretic level, a natural next step is to investigate whether the corresponding quantised theories admit distinctive, potentially testable signatures. This also raises a structural question: whether symplectic reduction and quantisation commute in general, or whether obstructions can arise in more complex systems.

More broadly, the present work isolates a baseline phenomenon: before invoking gravity, the interplay between Hamiltonian integrability, spectral structure, and finite-time observational resolution already constrains how many canonical directions are operationally relevant. Understanding how gravitational dynamics modifies, stabilises, or universalises such compression mechanisms remains an open and intriguing direction.

\begin{acknowledgments}
 V.K.\ gratefully acknowledges support from the \emph{Deutscher Akademischer Austauschdienst} (DAAD, German Academic Exchange Service) and thanks ChunJun Cao and Aditya Dwarkesh for insightful discussions. O.F.\ was supported by a Fraunhofer-Schwarzschild-Fellowship at Universit\"atssternwarte M\"unchen (LMU observatory) and by DFG's Excellence Cluster ORIGINS (EXC-2094 – 390783311). K.G.\ is grateful for the hospitality of the Perimeter Institute, where part of this work was carried out. Research at the Perimeter Institute is supported in part by the Government of Canada through the Department of Innovation, Science and Economic Development, and by the Province of Ontario through the Ministry of Colleges and Universities. This work was supported by a grant from the Simons Foundation (Grant No.~1034867, Dittrich). The authors also acknowledge the contribution of the COST Action CA23130, \emph{``Bridging high and low energies in search of quantum gravity (BridgeQG).''} We further thank the developers of the open-source Python libraries \verb|NumPy|~\cite{NumPy}, \verb|CuPy|~\cite{Okuta2017CuPyA}, and \verb|Matplotlib|~\cite{Matplotlib}, whose tools were indispensable for the numerical computations and visualizations presented in this work.
\end{acknowledgments}

\section*{Data availability}

The \verb|Python| code used to generate all numerical results and figures in this work is publicly available at \url{https://github.com/ScaleOfVarun/area-scaling-dynamical-dofs}. All data can be regenerated using the provided scripts.

\bibliographystyle{unsrtnat}
\bibliography{ClassHolography}

\appendix
\renewcommand{\thesection}{\Alph{section}}   
\renewcommand{\theHsection}{\Alph{section}}   


\section{Background on symplectic model order reduction}
\label{app:SMOR-background}

This appendix gives the geometric background behind the operational construction used in Sec.~\ref{sec:SMOR}. The main text only uses the finite-dimensional, linear version of SMOR, because that is sufficient for the regulated scalar field theory. Here we briefly explain the more general Hamiltonian setting in order to clarify why a symplectic reduction leads again to an autonomous Hamiltonian system.

\subsection{Hamiltonian systems and symplectic approximation}
\label{app:SympApprox}

The phase space of an unconstrained Hamiltonian system is a symplectic manifold $(\mathcal M,\omega)$. Here $\mathcal M$ is an even-dimensional smooth manifold and $\omega$ is a closed and nondegenerate two-form,
\begin{align}
    \dd\omega=0.
\end{align}
Given a Hamiltonian function $\mathcal H:\mathcal M\to\mathbb R$, the Hamiltonian vector field $X_{\mathcal H}$ is defined by
\begin{align}
    \iota_{X_{\mathcal H}}\omega = \dd\mathcal H,
    \label{eq:app-HamVF-def}
\end{align}
where $\iota_X$ denotes contraction with the vector field $X$. This equation is the coordinate-free form of Hamilton's equations. A trajectory is a curve $\bm z(t)$ satisfying
\begin{align}
    \dot{\bm z}(t)  = X_{\mathcal H}(\bm z(t)).
\end{align}
The associated Hamiltonian flow is denoted by $\Phi^t$, so that
\begin{align}
    \bm z(t)  = \Phi^t(\bm z_0).
\end{align}
In canonical Darboux coordinates
\begin{align}
    \bm z=(q^1,\ldots,q^N,p_1,\ldots,p_N),
\end{align}
the symplectic form is
\begin{align}
    \omega  = \dd q^j\wedge \dd p_j,
\end{align}
and Hamilton's equations become
\begin{align}
    \dot{\bm z} = \mathbb J_{2N}\nabla\mathcal H(\bm z).
\end{align}

In the applications of this paper, the infrared- and ultraviolet-regularised scalar field has precisely this canonical form. The full phase space is
\begin{align}
    \mathcal M  = \mathbb R^{2N},
\end{align}
with canonical coordinates
\begin{align}
    (Q^1,\ldots,Q^N,P_1,\ldots,P_N).
\end{align}
For each choice of physical parameters $\mu$, such as mass, region size, ultraviolet cutoff, and boundary condition, we denote the corresponding Hamiltonian flow by $\Phi_\mu^t$.

Model reduction starts from the observation that the ambient phase space may be much larger than the portion actually explored by the dynamics. If one varies parameters, initial conditions, and time, the set of states reached by the system is
\begin{align}
    \mathcal S_{\mathrm{sol}} := \big\{ \Phi_\mu^t(\bm z_0) \ \big|\ \mu\in\mathcal P,\; \bm z_0\in\mathcal Z,\;  t\in\mathbb R \big\} \subset\mathcal M,
    \label{eq:app-solution-set}
\end{align}
where $\mathcal P$ is the parameter domain and $\mathcal Z\subset\mathcal M$ is a set of admissible initial data. This object is often called the solution manifold in the model-reduction literature. The terminology is conventional: $\mathcal S_{\mathrm{sol}}$ need not be a smooth embedded submanifold of $\mathcal M$. For example, different solution branches may intersect, trajectories may densely fill invariant tori, and the dependence on parameters may fail to be smooth \cite{doi:10.1137/1.9781611974829.ch2,buchfink2024modelreductionmanifoldsdifferential}.

For Hamiltonian systems, the reduced space should not be just any low-dimensional approximation to $\mathcal S_{\mathrm{sol}}$. It should itself carry a symplectic structure inherited from the full phase space. Let $(\mathcal V,\widetilde\omega)$ be a $2m$-dimensional symplectic manifold with $m\ll N$. A smooth map
\begin{align}
    \xi:\mathcal V\hookrightarrow \mathcal M
\end{align}
is called a symplectic embedding if it is an injective immersion and preserves the symplectic form,
\begin{align}
    \xi^\ast\omega = \widetilde\omega.
    \label{eq:app-symplectic-embedding}
\end{align}
The image $\xi(\mathcal V)$ is then a symplectic submanifold of the full phase space \cite[Lemma~5.13]{buchfink2024modelreductionmanifoldsdifferential}. This is the geometric reason that reduced coordinates can still be interpreted as canonical variables.

In an abstract setting, the quality of the approximation can be measured by equipping $\mathcal M$ with a compatible distance $d_{\mathcal M}$ and considering the directed discrepancy
\begin{align}
    d_{\mathcal M}\big(  \mathcal S_{\mathrm{sol}},\xi(\mathcal V) \big) := \sup_{\bm x\in\mathcal S_{\mathrm{sol}}} \inf_{\bm y\in\mathcal V} d_{\mathcal M}\big(\bm x,\xi(\bm y)\big).
    \label{eq:app-set-gap}
\end{align}
In the linear finite-dimensional setting of the main text this abstract distance is replaced by the concrete Frobenius projection error of the snapshot matrix, Eq.~\eqref{eq:proj-loss}.

The present paper uses a trajectory-wise version of this idea. We do not seek one reduced space that represents a large family of solutions. Instead, for fixed parameters and initial data, we consider the single orbit
\begin{align}
    \Gamma_\mu(\bm z_0)  := \big\{  \Phi_\mu^t(\bm z_0)  \;:\;  t\in\mathbb R \big\}.
    \label{eq:app-single-orbit}
\end{align}
The question is how many canonical directions this orbit explores. The snapshot construction in Sec.~\ref{sec:snapshot-projection-error} is the finite-data implementation of this trajectory-wise question.

\subsection{Reduction maps and autonomous reduced Hamiltonian systems}
\label{app:ROM-background}

A low-dimensional symplectic subspace is only the first part of a reduced model. To obtain an autonomous reduced Hamiltonian system, one must also specify how full phase-space states and tangent vectors are mapped back to the reduced space. This is the role of the reduction map.

Let $(\mathcal M,\omega,\mathcal H)$ be a Hamiltonian system and let $(\mathcal V,\widetilde\omega)$ be a $2m$-dimensional reduced symplectic space. Suppose that
\begin{align}
    \xi:\mathcal V\hookrightarrow\mathcal M
\end{align}
is a symplectic embedding. A reduction map is constructed from a smooth left inverse
\begin{align}
    \varrho:\mathcal M\to\mathcal V, \qquad  \varrho\circ\xi   =  \operatorname{id}_{\mathcal V},
    \label{eq:app-left-inverse}
\end{align}
defined at least in a neighbourhood of $\xi(\mathcal V)$. The map $\varrho$ sends a full state to reduced coordinates. Its differential sends ambient tangent vectors to tangent vectors in the reduced space. Together they define
\begin{align}
    R:T\mathcal M\to T\mathcal V,  \qquad  (\bm z,v)  \mapsto \big( \varrho(\bm z),   \varrho_{\ast}|_{\bm z}(v) \big).
    \label{eq:app-reduction-map}
\end{align}
Compatibility with the embedding means
\begin{align}
    \varrho\circ\xi  =  \operatorname{id}_{\mathcal V},  \qquad  \varrho_\ast|_{\xi(\widetilde{\bm z})} \circ \xi_\ast|_{\widetilde{\bm z}} = \operatorname{id}_{T_{\widetilde{\bm z}}\mathcal V}.
    \label{eq:app-projection-property}
\end{align}

The reduced vector field is obtained by restricting the full Hamiltonian vector field to the embedded reduced space and projecting it back to $\mathcal V$:
\begin{align}
    X_{\widetilde{\mathcal H}}(\widetilde{\bm z})  := \varrho_\ast|_{\xi(\widetilde{\bm z})} \left[    X_{\mathcal H}(\xi(\widetilde{\bm z})) \right].
    \label{eq:app-reduced-vector-field}
\end{align}
The reduced-order model is then
\begin{align}
    \dot{\widetilde{\bm z}}(t) = X_{\widetilde{\mathcal H}}(\widetilde{\bm z}(t)), \qquad \widetilde{\bm z}(t_0) = \varrho(\bm z_0).
    \label{eq:app-ROM-general}
\end{align}
Because the embedding is symplectic, the reduced Hamiltonian is simply the pullback of the full Hamiltonian,
\begin{align}
    \widetilde{\mathcal H} = \xi^\ast\mathcal H= \mathcal H\circ\xi, \qquad \iota_{X_{\widetilde{\mathcal H}}}\widetilde\omega =    \dd\widetilde{\mathcal H}.
    \label{eq:app-pullback-Hamiltonian}
\end{align}
This is the main structural advantage of SMOR: the reduced equations are not merely a projected set of equations, but a Hamiltonian system on the reduced phase space.

This statement should not be confused with automatic preservation of all conserved quantities of the full system. Symplectic reduction preserves the Hamiltonian character of the dynamics and structural features tied to the symplectic form. Additional conserved quantities are inherited only when the corresponding symmetries or invariant structures are compatible with the embedded reduced space \cite{Afkham_2017,hesthaven2021structurepreservingmodelorderreduction}.

\subsection{Linear symplectic embeddings and the SMG map}
\label{app:linear-SMG-background}

The construction used in the main text is the linear version of the preceding geometric setup. The full phase space is $\mathcal M=\mathbb R^{2N}$ and the reduced phase space is $\mathcal V=\mathbb R^{2m}$, both equipped with their canonical symplectic forms. A linear symplectic embedding is represented by a matrix
\begin{align}
    \bm V\in\mathbb R^{2N\times 2m}, \qquad \bm V^\top\mathbb J_{2N}\bm V = \mathbb J_{2m}.
    \label{eq:app-linear-symplectic-basis}
\end{align}
The embedding is
\begin{align}
    \xi(\widetilde{\bm z})  = \bm V\widetilde{\bm z}.
\end{align}
A standard reduction map in this setting is the symplectic-manifold Galerkin map \cite{hesthaven2021structurepreservingmodelorderreduction,buchfink2024modelreductionmanifoldsdifferential}.  It uses the symplectic adjoint
\begin{align}
    \bm V^+  = \mathbb J_{2m}^{\top}\bm V^\top\mathbb J_{2N},
    \label{eq:app-symplectic-adjoint}
\end{align}
which satisfies
\begin{align}
    \bm V^+\bm V = \mathbb I_{2m}.
\end{align}
The associated projector is
\begin{align}
    P_{\bm V} = \bm V\bm V^+.
    \label{eq:app-symplectic-projector}
\end{align}
The reduced Hamiltonian and reduced equations are then those given in Eqs.~\eqref{eq:reduced-hamiltonian} and \eqref{eq:ROM}.

Thus the practical workflow used in Sec.~\ref{sec:SMOR} can be summarised as follows. One samples a full Hamiltonian trajectory and constructs a snapshot matrix. One then chooses a low-dimensional symplectic basis that minimises the projection error on these snapshots. This basis defines both a reconstruction map and an autonomous reduced Hamiltonian system. In this paper the basis is constructed using the complex-SVD/PSD procedure described in Sec.~\ref{sec:csvd-psd-basis}, and the resulting singular values determine the minimal symplectic dimension introduced in Sec.~\ref{sec:rank=dim}.

\section{Analytic construction of the embedding matrix \texorpdfstring{$\bm V^{(\rm opt)}$}{}}
\label{app:analytic-Vstar}

In this appendix we construct, \emph{analytically and in closed form}, the time--independent \emph{ortho--symplectic} embedding matrix $\bm V^{(\rm opt)}$ that maps reduced coordinates to the full phase space and exactly reproduces the trajectory generated by a free Hamiltonian field. The construction clarifies three key points that underlie the numerical results of the main text: (i) each \emph{distinct physical frequency} contributes exactly two independent time dependences (cosine and sine); (ii) degeneracies of the Laplacian spectrum affect only the \emph{orientation} of the reduced subspace, not its dimension; and (iii) the reduced symplectic subspace selected by SMOR coincides with an \emph{invariant} symplectic subspace of the Hamiltonian flow. We also discuss a natural sampling normalisation under which the construction stabilises in the long--time limit.

\subsection{Snapshot factorisation by frequency shells}

We consider a free Hamiltonian system consisting of $N$ independent harmonic oscillators, labelled by $\alpha$, with frequencies $\Omega_\alpha$ and canonical coordinates $(Q_\alpha,P_\alpha)$,
\begin{align}
  \mathcal{H}(\bm Q,\bm P)
  = \tfrac12\,\bm P^\top \bm P
    + \tfrac12\,\bm Q^\top \Omega^2 \bm Q,
  \qquad
  \Omega^2 := \mathrm{diag}(\Omega_\alpha^2).
  \label{eq:app-H}
\end{align}
Hamilton’s equations are solved exactly by
\begin{align}
  Q_\alpha(t)
  &= A_\alpha \cos(\Omega_\alpha t)
     + B_\alpha \sin(\Omega_\alpha t), \nonumber\\
  P_\alpha(t)
  &= -A_\alpha\Omega_\alpha \sin(\Omega_\alpha t)
     + B_\alpha\Omega_\alpha \cos(\Omega_\alpha t),
  \label{eq:app-sol}
\end{align}
where the real amplitudes $(A_\alpha,B_\alpha)$ are fixed by the initial condition.

To analyse the structure of the trajectory in phase space, we introduce the complex combination
\begin{align}
  X_\alpha(t) := Q_\alpha(t) + i\,P_\alpha(t),
\end{align}
and collect its values at sampling times $\{t_j\}_{j=1}^T$ into the complex snapshot matrix
\begin{align}
  \mathbb X_c \in \mathbb C^{N\times T},
  \qquad
  (\mathbb X_c)_{\alpha j}
  := X_\alpha(t_j).
  \label{eq:app-Xc}
\end{align}
As discussed in the main text, the left singular vectors of $\mathbb X_c$ determine the symplectic subspace selected by SMOR.

A crucial simplification arises by grouping modes into \emph{frequency shells}.  Modes with identical (or numerically indistinguishable) frequencies are assigned to the same shell. Let $S$ denote the set of distinct shells, labelled by $s$, with representative frequency $\Omega_s$. For each shell we define the index set $\mathcal I_s\subset\{1,\ldots,N\}$ and its degeneracy $g_s := |\mathcal I_s|$. After reordering rows by shells, the snapshot matrix decomposes as
\begin{align}
  \mathbb X_c
  =
  \begin{bmatrix}
    \mathbb X_c^{(1)} \\
    \vdots \\
    \mathbb X_c^{(n_\Omega)}
  \end{bmatrix},
  \qquad
  \mathbb X_c^{(s)} \in \mathbb C^{g_s\times T},
\end{align}
where $n_\Omega$ is the number of distinct frequencies.

For a fixed shell $s$, all modes share the same temporal dependence. We therefore introduce the universal time vectors
\begin{align}
  c^{(s)} &:= (\cos(\Omega_s t_1),\ldots,\cos(\Omega_s t_T))^\top, \\
  s^{(s)} &:= (\sin(\Omega_s t_1),\ldots,\sin(\Omega_s t_T))^\top,
\end{align}
and shell--dependent complex spatial coefficients
\begin{align}
  a^{(+)}_\alpha &:= A_\alpha + i\,\Omega_s B_\alpha, &
  a^{(-)}_\alpha &:= B_\alpha - i\,\Omega_s A_\alpha,
  \qquad \alpha\in\mathcal I_s .
\end{align}
Collecting these into matrices,
\begin{align}
  A^{(s)}
  := \begin{bmatrix} a^{(+)} & a^{(-)} \end{bmatrix}
  \in \mathbb C^{g_s\times 2},
  \qquad
  T^{(s)}
  := \begin{bmatrix} c^{(s)} & s^{(s)} \end{bmatrix}
  \in \mathbb R^{T\times 2},
\end{align}
a direct substitution of Eq.~\eqref{eq:app-sol} into
Eq.~\eqref{eq:app-Xc} yields the exact factorisation
\begin{align}
  \mathbb X_c^{(s)} = A^{(s)}\,T^{(s)\top}.
  \label{eq:shell-factor}
\end{align}
Equation~\eqref{eq:shell-factor} makes explicit that each excited frequency shell contributes at most two independent time harmonics (cosine and sine), irrespective of the degeneracy $g_s$. All dependence on the initial amplitudes and phases resides in the two columns of $A^{(s)}$.

\subsection{Shell-wise singular structure and reduced dimension}

To extract the singular structure of $\mathbb X_c^{(s)}$, we introduce the time and amplitude Gram matrices
\begin{align}
  \mathbb M_t^{(s)}
  &:= T^{(s)\top}T^{(s)}
   = \begin{bmatrix}
       t_{cc} & t_{cs} \\
       t_{cs} & t_{ss}
     \end{bmatrix}
     \in \mathbb R^{2\times2}, \\
  \mathbb M_a^{(s)}
  &:= A^{(s)\dagger}A^{(s)}
   = \begin{bmatrix}
       a_{++} & a_{+-} \\
       a_{-+} & a_{--}
     \end{bmatrix}
     \in \mathbb C^{2\times2}.
\end{align}
Using Eq.~\eqref{eq:shell-factor} one finds
\begin{align}
  \mathbb X_c^{(s)}\mathbb X_c^{(s)\dagger}
  = A^{(s)}\,\mathbb M_t^{(s)}\,A^{(s)\dagger}.
\end{align}
Hence the two nonzero singular values of $\mathbb X_c^{(s)}$ are given by the eigenvalues of the $2\times2$ matrix
$\mathbb M_t^{(s)}\mathbb M_a^{(s)}$,
\begin{align}
  \sigma_{s,\pm}^2
  \in \mathrm{spec}\!\big(\mathbb M_t^{(s)}\mathbb M_a^{(s)}\big).
  \label{eq:sv-2x2}
\end{align}
Explicitly,
\begin{align}
  \sigma_{s,\pm}^2
  =
  \frac12\!\left[
    \mathrm{tr}(\mathbb M_t^{(s)}\mathbb M_a^{(s)})
    \pm
    \sqrt{
      \mathrm{tr}^2(\mathbb M_t^{(s)}\mathbb M_a^{(s)})
      - 4\,\det(\mathbb M_t^{(s)})\det(\mathbb M_a^{(s)})
    }
  \right].
\end{align}

Let $z_\pm\in\mathbb C^2$ denote the corresponding normalised eigenvectors. The left singular vectors are then
\begin{align}
  \tilde u_{s,\pm} := A^{(s)} z_\pm \in \mathbb C^{g_s},
  \qquad
  u_{s,\pm}
  := \frac{\tilde u_{s,\pm}}{\|\tilde u_{s,\pm}\|_2}.
\end{align}
Stacking all shells yields the full analytic left--singular matrix
\begin{align}
  \mathbb U_{\mathrm{an}}
  :=
  \big[\,u_{s,+}\ \ u_{s,-}\,\big]_{s\in S}
  \in \mathbb C^{N\times r},
  \qquad
  r = \sum_{s\in S} r_s,
  \quad r_s\in\{1,2\}.
\end{align}
Generically $r_s=2$ for each excited shell, so that $r=2n_\Omega$ and the real reduced phase--space dimension is $2r=4n_\Omega$.

\subsection{Real ortho--symplectic lift and geometric interpretation}

To obtain a real symplectic embedding, we write $\mathbb U_{\mathrm{an}} = \bm\Phi + i\,\bm\Psi$ with $\bm\Phi,\bm\Psi\in\mathbb R^{N\times r}$. Orthonormality of $\mathbb U_{\mathrm{an}}$ implies
\begin{align}
  \bm\Phi^\top\bm\Phi + \bm\Psi^\top\bm\Psi &= \mathbb I_r,
  &
  \bm\Phi^\top\bm\Psi &= \bm\Psi^\top\bm\Phi .
\end{align}
We then define the real embedding matrix
\begin{align}
  \bm V^\star
  :=
  \begin{bmatrix}
    \bm\Phi & -\bm\Psi \\
    \bm\Psi & \ \bm\Phi
  \end{bmatrix}
  \in \mathbb R^{2N\times2r},
  \label{eq:Vstar-def}
\end{align}
which satisfies
\begin{align}
  {\bm V^{(\rm opt)}}^\top\mathbb J_{2N}\bm V^{(\rm opt)} = \mathbb J_{2r},
  \qquad
  {\bm V^{(\rm opt)}}^\top\bm V^{(\rm opt)}= \mathbb I_{2r}.
\end{align}
Thus $\bm V^{(\rm opt)}$ is ortho--symplectic and provides the desired embedding.

Geometrically, each shell $s$ contributes a four--dimensional symplectic subspace $\mathcal V_s$ (two canonical pairs) corresponding to the cosine and sine time dependences at frequency $\Omega_s$. Within the $g_s$--dimensional degenerate eigenspace, the embedding selects the particular two--plane determined by the initial amplitudes. The full reduced phase space is the symplectic direct sum
\begin{align}
  \mathcal V
  = \bigoplus_{s\in S} \mathcal V_s,
  \qquad
  \dim\mathcal V_s = 4,
\end{align}
with the restricted symplectic form
\begin{align}
  \omega|_{\mathcal V}
  =
  \sum_{s\in S}
  \left(
    d\widetilde q_{s,c}\wedge d\widetilde p_{s,c}
    + d\widetilde q_{s,s}\wedge d\widetilde p_{s,s}
  \right).
\end{align}

\subsection{Sampling normalisation and numerical validation}

With the raw Euclidean time inner product, the entries of $\mathbb M_t^{(s)}$ scale linearly with the number of samples $T$. To remove this trivial dependence we rescale the snapshots by $1/\sqrt T$, equivalently using the time--averaged inner product
\begin{align}
  \langle u,v\rangle_t
  := \frac{1}{T}\sum_{j=1}^T \overline{u(t_j)}\,v(t_j).
\end{align}
For uniformly spaced samples over a long observation window and nonresonant sampling one finds
\begin{align}
  \frac{1}{T}\,\mathbb M_t^{(s)} \longrightarrow \tfrac12\,\mathbb I_2\quad \text{and} \qquad
  \frac{1}{T}\,\langle c^{(s)},s^{(s)}\rangle \to 0,
\end{align}
so that the singular values converge to constants determined solely by $\mathbb M_a^{(s)}$.

\begin{figure}[t]
    \centering
    \begin{subfigure}{0.48\textwidth}
        \includegraphics[width=\linewidth]{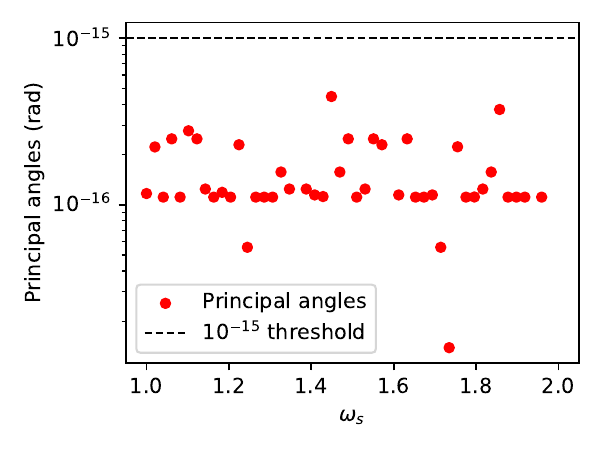}
        \caption{Principal angles between analytic and numerical singular subspaces.}
    \end{subfigure}
    \hfill
    \begin{subfigure}{0.48\textwidth}
        \includegraphics[width=\linewidth]{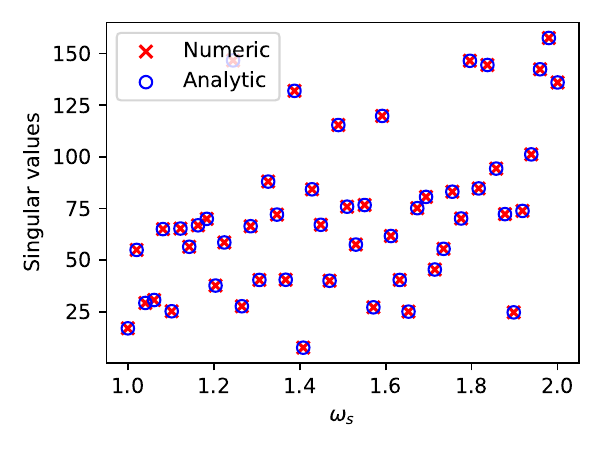}
        \caption{Singular value comparison (analytic vs.\ numerical).}
    \end{subfigure}
    \caption{
        Validation of the analytic construction of the symplectic basis $\bm V^{(\rm opt)}$.
        Panel~(a) shows the principal angles between the analytic subspace spanned by
        $\mathbb{U}_{\mathrm{an}}$ and the numerical left–singular subspace obtained by direct SVD
        of the complex snapshot matrix $\mathbb{X}_c$. 
        All angles are smaller than $10^{-15}\,\mathrm{rad}$, demonstrating that the analytic and
        numerical subspaces are indistinguishable to machine precision. 
        Panel~(b) compares the singular values $\sigma_{s,\pm}$ obtained analytically from
        Eq.~\eqref{eq:sv-2x2} with those computed numerically by SVD.
        The two sets coincide to within relative differences below $10^{-15}$ in all cases. 
        Together, these results confirm that the analytic expressions for the singular vectors and singular
        values---derived solely from the initial amplitudes and mode frequencies---exactly reproduce the numerical outcome. 
        The symplectic basis $\bm V^{(\rm opt)}$ can thus be obtained fully analytically, without any numerical decomposition.
    }
    \label{fig:analytic_vs_numeric}
\end{figure}

Finally, we compare the analytic singular vectors and singular values with a direct numerical SVD of $\mathbb X_c$. The principal angles between the analytic and numerical subspaces are below $10^{-15}\,$rad and the relative singular--value errors are below $10^{-15}$ in all cases (Fig.~\ref{fig:analytic_vs_numeric}), confirming that the analytic construction reproduces the numerical SMOR basis to machine precision.

\section{Numerical setup for projection--error analysis}
\label{app:numsetup}

This appendix summarises the numerical configuration used to produce the projection--error plots in Figs.~\ref{fig:projection_error_freefield} and~\ref{fig:projection_error_lambda}. The full implementation, including all scripts used to generate the figures in the main text and appendices, is available in the repository at \url{https://github.com/ScaleOfVarun/area-scaling-dynamical-dofs}.

\subsection{Field theory and discretisation}

We evolve a real scalar field with quartic self--interaction on a periodic cubic box $\mathcal B=[-L_{\rm IR}/2,L_{\rm IR}/2]^3$ governed by the Hamiltonian
\begin{align}
  \mathcal{H}[\phi,\pi]
  = \int_{\mathcal{B}}\!\mathrm{d}^3x
    \bigg[
      \frac{1}{2}\,\pi^2
      + \frac{1}{2}(\nabla\phi)^2
      + \frac{1}{2}m^2\phi^2
      + \frac{\lambda}{4}\phi^4
    \bigg],
  \qquad m=1 .
\end{align}
We work in dimensionless units with $c=\hbar=1$, so $m$ sets the natural time unit. All times reported below are expressed in these units.

The field is discretised on a uniform lattice with
\begin{align}
  L_{\rm IR}=20,
  \qquad
  N^3 = 14^3 \ \text{grid points},
\end{align}
and periodic boundary conditions. Spatial derivatives are evaluated spectrally using FFTs, so that in Fourier space $-\Delta\phi \mapsto k^2\phi_{\bm k}$. The nonlinear term is evaluated pointwise in configuration space. For the parameter range considered, aliasing effects do not visibly affect the diagnostics reported here.

\subsection{Time integration, observation window, and frequency resolution}

Time evolution is performed with the velocity--Verlet scheme applied to the canonical variables $(Q(t), P(t))$.

For the free theory, the timestep is chosen from a linear stability bound. Let
\begin{align}
  \Omega_{\max} = \max_{\bm k}\sqrt{m^2+|\bm k|^2},
\end{align}
then we set
\begin{align}
  \Delta t_0 = \frac{0.2}{\Omega_{\max}},
\end{align}
choose an integer number of steps $N_{\rm step}=\lceil T_{\rm obs}/\Delta t_0\rceil$, and finally define $\Delta t = T_{\rm obs}/N_{\rm step}$ so that the final time lands exactly on $T_{\rm obs}$. For interacting runs we use the same procedure but with a slightly stricter nonlinear safety factor, as implemented in the code.

All projection--error results shown in the main text are computed from trajectories sampled over the observation window
\begin{align}
  T_{\rm obs} = 14\,L_{\rm IR} = 280.
\end{align}
This choice implies a finite temporal frequency resolution
\begin{align}
  \Delta\Omega \sim \frac{\pi}{T_{\rm obs}} = \frac{\pi}{280},
\end{align}
which sets the scale below which distinct frequencies cannot be operationally resolved from finite--time data.

\subsection{Initial conditions}

\paragraph{Free theory ($\lambda=0$).}
For Fig.~\ref{fig:projection_error_freefield} we use generic Gaussian random initial data in real space,
\begin{align}
  \phi(\bm x,0),\ \pi(\bm x,0) \sim \mathcal N(0,\sigma^2),
  \qquad \sigma = 10^{-2},
\end{align}
drawn independently at each lattice site. We evolve $n_{\rm ic}=3$ independent realisations (fixed RNG seed in the code) to illustrate robustness with respect to initial phases and amplitudes.

\paragraph{Weakly interacting theory ($\lambda\phi^4$).}
For Fig.~\ref{fig:projection_error_lambda} the initial amplitudes are chosen to mimic free quantum vacuum fluctuations mode--by--mode: in Fourier space, each mode is drawn from a complex Gaussian distribution with variances
\begin{align}
  \mathrm{Var}(q_{\bm k})=\frac{1}{2\Omega_{\bm k}},
  \qquad
  \mathrm{Var}(p_{\bm k})=\frac{\Omega_{\bm k}}{2},
  \qquad
  \Omega_{\bm k}=\sqrt{m^2+|\bm k|^2},
\end{align}
with the appropriate reality constraints for a real field (implemented via an rFFT representation). We fix one such initial realisation and vary the coupling over
\begin{align}
  \lambda \in \{\,0,\ 10^{-6},\ 10^{-4},\ 10^{-2}\,\},
\end{align}
remaining in the weakly nonlinear regime as quantified in App.~\ref{app:weak-nonlinearity-diagnostics}.

\subsection{Snapshot matrix and projection error}

From the saved snapshots we construct, for each trajectory, the complex snapshot matrix
\begin{align}
  \mathbb X_c
  \in \mathbb C^{N_{\rm dof}\times T_{\rm snap}},
  \qquad
  (\mathbb X_c)_{ij}=Q_i(t_j)+i\,P_i(t_j),
\end{align}
where $i$ indexes lattice sites ($N_{\rm dof}=N^3$) and $t_j$ are sampling times. The relative projection error is computed from the eigenvalues of the Gram matrix $G=\mathbb X_c^\dagger\mathbb X_c$: if $\{s_\alpha\}$ are the singular values of $\mathbb X_c$, then
\begin{align}
  \mathcal E_{\mathrm{proj}}(r)
  =
  \frac{\big(\sum_{\alpha>r} s_\alpha^2\big)^{1/2}}{\big(\sum_{\alpha\ge1} s_\alpha^2\big)^{1/2}},
\end{align}
which is the quantity plotted in Figs.~\ref{fig:projection_error_freefield} and~\ref{fig:projection_error_lambda} as a function of the reduced dimension $2r$.

The vertical dashed line in both plots corresponds to the free--theory prediction
\begin{align}
  d_{\rm theory}=4\,n_\Omega(T_{\rm obs}),
\end{align}
where $n_\Omega(T_{\rm obs})$ denotes the number of \emph{operationally distinct} free--field frequencies $\Omega_{\bm k}$ below the UV cutoff when frequencies are identified up to the finite resolution $|\Delta\Omega|\lesssim \pi/T_{\rm obs}$, as implemented in the code. This is the appropriate benchmark for finite--time snapshot data.

\section{Diagnostics for weak nonlinearity}
\label{app:weak-nonlinearity-diagnostics}

This appendix documents two simple checks that support the claim made in Sec.~\ref{sec:weak-interactions}: for small coupling and small amplitudes, the dynamics stay in the quasi–integrable regime over the observation window, so the number of basic frequencies that actually drive the motion (and hence the minimal reduced dimension) is essentially unchanged.

\subsection{A small nonlinearity budget \texorpdfstring{$\varepsilon(t)$}{epsilon(t)}}

\begin{figure}[t]
  \centering
  \includegraphics[width=0.65\textwidth]{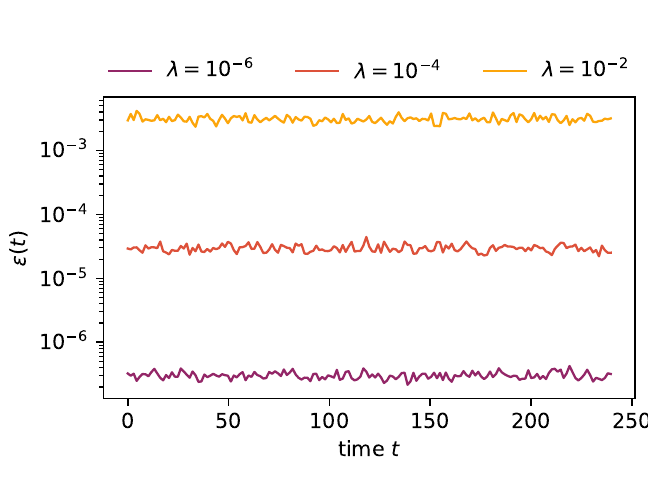}
  \caption{Nonlinearity budget $\varepsilon(t)$
  as defined in~\eqref{eq:eps-appendix} for several $\lambda>0$.
  Curves remain well below unity over the entire time interval, indicating that
  the quartic term acts as a small perturbation of the quadratic dynamics.}
  \label{fig:epsilon-smallness}
\end{figure}

We quantify “weakly interacting” by monitoring the dimensionless ratio between quartic and quadratic energies,
\begin{align}
  \varepsilon(t)
  \;:=\;
  \frac{\lambda \int_{\mathcal B}\phi(t,x)^4\,\mathrm{d}^3x}
       {\int_{\mathcal B}\!\Big(P(t,x)^2 + Q(t,x)\,(m^2-\nabla^2)\,Q(t,x)\Big)\,\mathrm{d}^3x}.
  \label{eq:eps-appendix}
\end{align}
On our finite box this is evaluated by straightforward Riemann sums on the grid; the Laplacian term is computed with the same discrete operator used in the time evolution. Fig.~\ref{fig:epsilon-smallness} shows $\varepsilon(t)$ for the couplings used in the main text (excluding $\lambda=0$). For all small couplings displayed, the curves remain $\ll 1$ throughout the entire observation window, confirming that the quartic energy stays a small correction to the quadratic energy. In particular, the regime assumed by the quasi–integrable picture (see Sec.~\ref{sec:weak-interactions}) is satisfied for these runs.

\subsection{Frequency confirmation map}

\begin{figure}[t]
  \centering
  \includegraphics[width=0.65\textwidth]{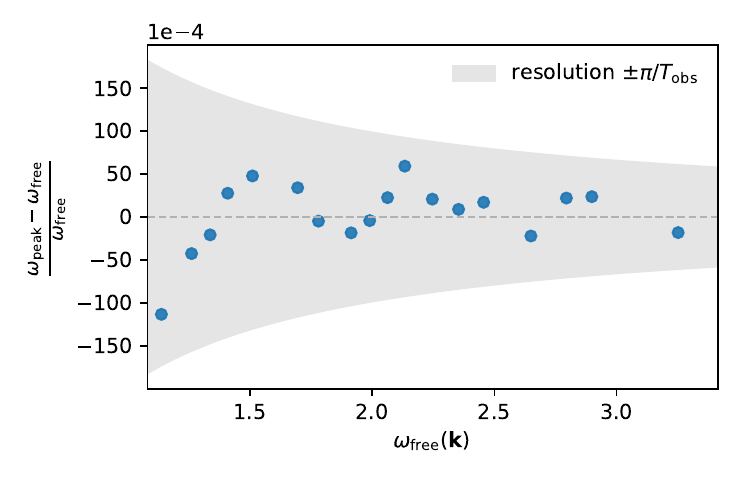}
  \caption{Fractional frequency shift
  $(\omega_{\mathrm{peak}}-\omega_{\mathrm{free}})/\omega_{\mathrm{free}}$
  versus $\omega_{\mathrm{free}}$ for a representative set of modes and several
  $\lambda$.
  The shaded band is the finite–time resolution $\sim 2\pi/T_{\mathrm{obs}}$
  expressed as a relative tolerance. Points within the band are
  unresolved at this time baseline; small excursions at larger $\lambda$
  reflect mild, expected frequency renormalisation.}
  \label{fig:frequency-map}
\end{figure}

As a complementary check, we verify that the observed peak frequencies remain close to the free values on the same time window. Pick a small, representative set of modes with distinct wave\-numbers (and hence distinct free frequencies $\omega_{\mathrm{free}}=\sqrt{m^2+|\mathbf k|^2}$). For each mode and coupling, we take the time series of the corresponding field pair $(Q(t),P(t))$ and form the complex signal $Q(t)+iP(t)$. We then compute its discrete Fourier transform in time over the full observation interval and identify the peak frequency $\omega_{\mathrm{peak}}$ of the power spectrum. The fractional shift,
\begin{align}
  \delta\omega / \omega_{\mathrm{free}}
  \;:=\;
  \frac{\omega_{\mathrm{peak}}-\omega_{\mathrm{free}}}{\omega_{\mathrm{free}}},
\end{align}
is plotted against $\omega_{\mathrm{free}}$ in Fig.~\ref{fig:frequency-map}. The shaded band indicates the finite–time frequency resolution inherited from the total duration $T_{\mathrm{obs}}$ (i.e.\ the smallest resolvable spacing $\sim 2\pi/T_{\mathrm{obs}}$ shown as a relative tolerance). Points inside the band are indistinguishable from zero shift at our resolution; points slightly outside indicate a small renormalisation consistent with weak nonlinear effects.

For the small couplings relevant to our study, almost all markers lie within (or only marginally outside) the resolution band. This confirms that, over the same $T_{\mathrm{obs}}$, the basic frequency content of the motion is essentially that of the free theory, in line with the persistence picture used in the main text.

Together, the smallness of $\varepsilon(t)$ and the near–identity of the frequency map show that, for our finite cutoff, small amplitudes, and small $\lambda$, the dynamics remain quasi–integrable over the observation window. Consequently, the number of independent basic frequencies that drive the motion does not proliferate, and the minimal real reduced dimension stays at $2r=4\,n_\Omega$ on the timescales considered.


\end{document}